\documentclass[fleqn,usenatbib]{mnras}
\usepackage{graphicx} 
\usepackage{amsmath}
\usepackage{mathtools}
\usepackage{newtxtext,newtxmath}
\usepackage{tabularx}
\usepackage{array}
\usepackage{url}
\usepackage{makecell}
\usepackage{tikz}
\usepackage[frozencache,cachedir=.]{minted}
\usepackage{listings}

\usepackage{orcidlink}
\usepackage{hyperref}
\hypersetup{
    colorlinks=true,
    linkcolor=blue,
    filecolor=magenta,      
    urlcolor=cyan,
    pdftitle={Overleaf Example},
    pdfpagemode=FullScreen,
    }

\title[Auriga discs and dwarfs]{Overview and public data release of the augmented Auriga Project: cosmological simulations of dwarf and Milky Way-mass galaxies}

\author[R. J. J. Grand et al.]{\parbox[t]{\textwidth}{%
Robert J. J. Grand \textsuperscript{\orcidlink{0000-0001-9667-1340}}$^1$\thanks{E-mail: r.j.grand@ljmu.ac.uk}, Francesca Fragkoudi \textsuperscript{\orcidlink{0000-0002-0897-3013}}$^2$, Facundo A. G{\'o}mez \textsuperscript{\orcidlink{0000-0003-4232-8584}}$^{3,4}$, Adrian Jenkins \textsuperscript{\orcidlink{0000-0003-4389-2232}}$^2$, Federico Marinacci \textsuperscript{\orcidlink{0000-0003-3816-7028}}$^5$, R{\"u}diger Pakmor \textsuperscript{\orcidlink{0000-0003-3308-2420}}$^6$ and Volker Springel \textsuperscript{\orcidlink{0000-0001-5976-4599}}$^6$}
\vspace{8pt}
\\%
$^1$Astrophysics Research Institute, Liverpool John Moores University, 146 Brownlow Hill, Liverpool, L3 5RF, UK\\%
$^2$Institute for Computational Cosmology, Department of Physics, Durham University, South Road, Durham, DH1 3LE, UK\\%
$^3$Instituto de Investigaci\'on Multidisciplinar en Ciencia y Tecnolog\'ia, Universidad de La Serena, Ra\'ul Bitr\'an 1305, La Serena, Chile\\%
$^4$Departamento de Astronom\'ia, Universidad de La Serena, Av. Juan Cisternas 1200 Norte, La Serena, Chile\\%
$^5$Department of Physics \& Astronomy ``Augusto Righi'', University of Bologna, via Gobetti 93/2, 40129 Bologna, Italy\\%
$^6$Max-Planck-Institut f\"{u}r Astrophysik, Karl-Schwarzschild-Str. 1, 85748 Garching, Germany
}

\date{Accepted XXX. Received YYY; in original form ZZZ}

\pubyear{2024}

\begin{document}

\maketitle

\begin{abstract}
We present an extended suite of the Auriga cosmological gravo-magnetohydrodynamical ``zoom-in'' simulations of 40 Milky Way-mass halos and 26 dwarf galaxy-mass halos run with the moving-mesh code {\sc arepo}. Auriga adopts the $\Lambda$ Cold Dark Matter ($\Lambda$CDM) cosmogony and includes a comprehensive galaxy formation physics model following the coupled cosmic evolution of dark matter, gas, stars, and supermassive black holes which has been shown to produce numerically well-converged galaxy properties for Milky Way-mass systems. We describe the first public data release of this augmented suite of Auriga simulations, which includes raw snapshots, group catalogues, merger trees, initial conditions, and supplementary data, as well as public analysis tools with worked examples of how to use the data. To demonstrate the value and robustness of the simulation predictions, we analyse a series of low-redshift global properties that compare well with many observed scaling relations, such as the Tully-Fisher relation, the star-forming main sequence, and HI gas fraction/disc thickness. Finally, we show that star-forming gas discs appear to build rotation and velocity dispersion rapidly for $z\gtrsim 3$ before they ``settle'' into ever-increasing rotation-dispersion ratios ($V/\sigma$). This evolution appears to be in rough agreement with some kinematic measurements from H$\alpha$ observations, and demonstrates an application of how to utilise the released data.
\end{abstract}


\section{Introduction}

Cosmological hydrodynamical simulations have become valuable and widely-used resources for the study of the formation and evolution of galaxies. These simulations begin from initial conditions set by the $\Lambda$ Cold Dark Matter ($\Lambda$CDM) cosmological model \citep{DEF85}, and include baryonic physics models \citep{KG91} that aim to evolve forward in time dark matter, gas, stars, black holes, and even magnetic fields. As such, they provide predictions and tests for both cosmology and complex physical processes operating across cosmic time on a wide range of scales. 

There are two main types of cosmological simulations discussed frequently in the literature. The first are large-box simulations that can resolve individual galaxies on kiloparsec scales within volumes of hundreds of megaparsecs; recent examples include Illustris \citep{VGS13}, EAGLE \citep{SCB15}, Horizon-AGN \citep{DPP16}, TNG \citep[e.g.][]{PNH18,NPS18,Naiman2018,Marinacci2018,Springel2018}, Simba \citep{Dave2019}, and FIREbox \citep{Feldmann2023}. These simulations produce predictions for a wide range of observable properties of the Cosmos, including the clustering of matter and the demographics of whole galaxy populations. The second type are cosmological ``zoom-in'' simulations, which enhance the resolution of stars, gas, and dark matter within and around individual systems and degrade the resolution of more distat matter in order to retain the large scale gravitational tidal field. Many groups around the world have adopted this technique to simulate the formation of halos across a wide range of masses. Particular interest is paid to low-mass galaxies \citep[including][]{WDS15,Revaz2018,Wheeler2019,Rey2019} and Milky Way-mass/Local Group systems including Eris \citep{GC11}, Aquarius \citep{MPS14}, APOSTLE \citep{FNS16}, Latte \citep{Wetzel+Hopkins+Kim+16}, Auriga \citep{GGM17}, FIRE \citep{Garrison-Kimmel2019}, NIHAO \citep{BOM20}, VINTERGATAN \citep{ARF20}, Hestia \citep{LCG20}, Justice League \citep{Applebaum_etal2021}, Artemis \citep{FMP20}, and EMP Pathfinder \citep{Reina-Campos2022}.

The wide range of numerical simulations now available pave the way to make powerful and timely interpretations of the wealth of observational data pertaining to galaxies and their stellar populations. In particular, ongoing and upcoming Galactic surveys are returning ever richer datasets through photometric/astrometric surveys like Gaia DR3 \citep{Gaia2016b,Gaiadr32023}, Rubin \citep{Rubin2019}, Roman, and spectroscopic surveys such as SDSS/APOGEE \citep{Majewski2017}, Gaia-ESO \citep{Gilmore2022}, 4MOST \citep{4most2019}, WEAVE \citep{weave2023}, and DESI \citep{Cooper2023}, which together provide detailed chemodynamical information for billions of individual stars. At the same time, IFU surveys such as CALIFA \citep{San12}, SAMI \citep{Cortese2014}, MANGA \citep{Bundy2015}, TIMER \citep{Gadotti2019}, PHANGS \citep{Emsellem2022}, and GECKOS \citep{vandesande2023} provide spatially-resolved information for external galaxies, and NASA's flagship JWST is providing new high redshift data that may require us to rethink several aspects of our understanding of galaxy formation models.

Each simulation suite (some of which are listed above) either uses different physics models (and implementations thereof), numerical methods (e.g. Lagrangian vs. Eulerian hydrodynamics), initial conditions, or a combination of these. This diversity results in a rich theoretical data bank that highlights uncertainty in observable predictions. Indeed, there have been efforts to understand the dependence of galaxy properties on different codes and physics models for the same set of initial conditions \citep[e.g. the Agora project:][]{Kim2014}. To help marginalise over these uncertainties and dependencies, ideally all available simulations should be studied, compared, and utilised in as many ways as possible. For these purposes, public dissemination of simulation data to the wider community is crucial \citep[e.g.][]{Nelson2019a,Wetzel2023}. In this spirit, this article describes the first public data release of the Auriga project -- a suite of 40 magneto-hydrodynamic cosmological zoom-in simulations of the formation of Milky Way-mass halos. A subset of these is available at higher resolution. In addition, we release data for two dwarf galaxy simulation suites, the first of which comprises 12 simulations of halos in the mass range of $[0.5, 5]\times 10^{11}$ $\rm M_{\odot}$ at $z=0$ (run at two different resolution levels), whereas the second comprises a suite of 14 simulations of halos in the mass range of $[0.5,5]\times 10^{10}$ $\rm M_{\odot}$ at $z=0$ (run at only the high resolution level). The data release includes raw snapshots containing information about the dark matter particles, star particles, gas cells, and black holes for each output time, group catalogues, merger trees, and supplementary data catalogues. We provide an overview of the simulations, and instructions on how to access and download the data. We release also a publicly available python-based analysis code and provide some worked examples of how to load and analyse the data and produce some basic plots. 

This paper gives an overview of the entire set of simulations. In Section~\ref{desc}, we provide a detailed description of the simulations, including: initial conditions and halo selection; main numerical methods employed; and the physics model. In Section~\ref{dataprod}, we thoroughly describe the released data products, and provide instructions on how to access and download the data. In Section~\ref{usingdata}, we discuss some considerations for using the data, and provide information on a publicly accessible python-based analysis package. In Section~\ref{science}, we present a first analysis of all simulations together in terms of several local observational scaling relations, in addition to the evolution of gas kinematics as an example of their predictive power for higher redshift observations. We summarise in Section~\ref{summary}.

\section{Description of the Simulations}
\label{desc}

Auriga is a suite of gravo-magnetohydrodynamic cosmological zoom-in simulations run with the moving-mesh code {\sc arepo}. All simulations follow the evolution of gas, dark matter, stars, and black holes according to a comprehensive galaxy formation model from a starting redshift of 127 down to the present-day. Our simulation suites include a set of 40 halos selected to be within a mass range of $0.5 < M_{200}/[10^{12}\, {\rm M_{\odot}}] < 2$ at redshift zero (thus covering the suspected halo mass of the Milky Way); 12 dwarf galaxy halos with masses $0.5 < M_{200}/[10^{11}\,{\rm M_{\odot}}] < 5$ and 14 low-mass dwarf galaxy halos with masses $0.5 < M_{200}/[10^{10}\,{\rm M_{\odot}}] < 5$. Here, $M_{200}$ is defined as the mass contained inside the radius, $R_{200}$, at which the mean enclosed mass volume density equals 200 times the critical density for closure.

\begin{figure*}
\includegraphics[scale=0.43,trim={0 0 1cm 0}, clip]{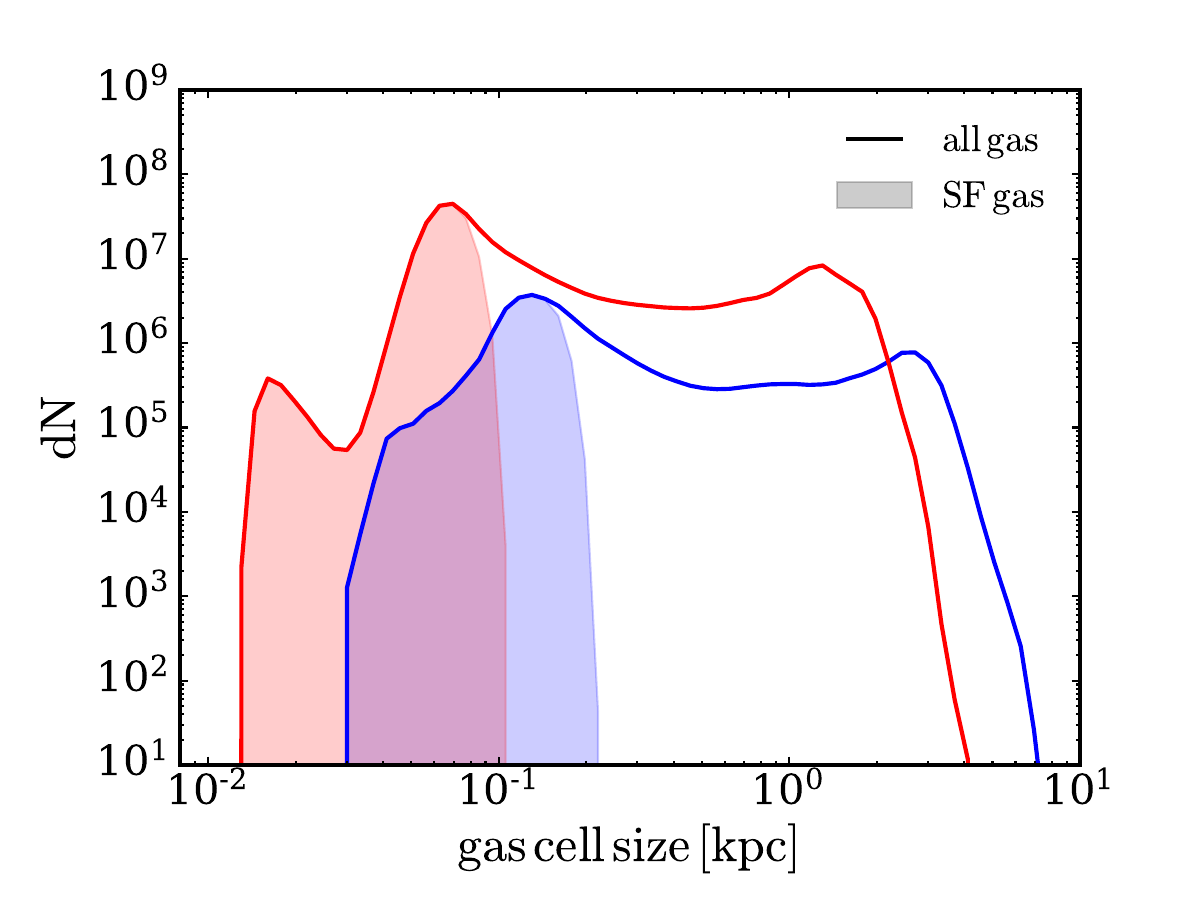}
\includegraphics[scale=0.43,trim={0 0 1cm 0}, clip]{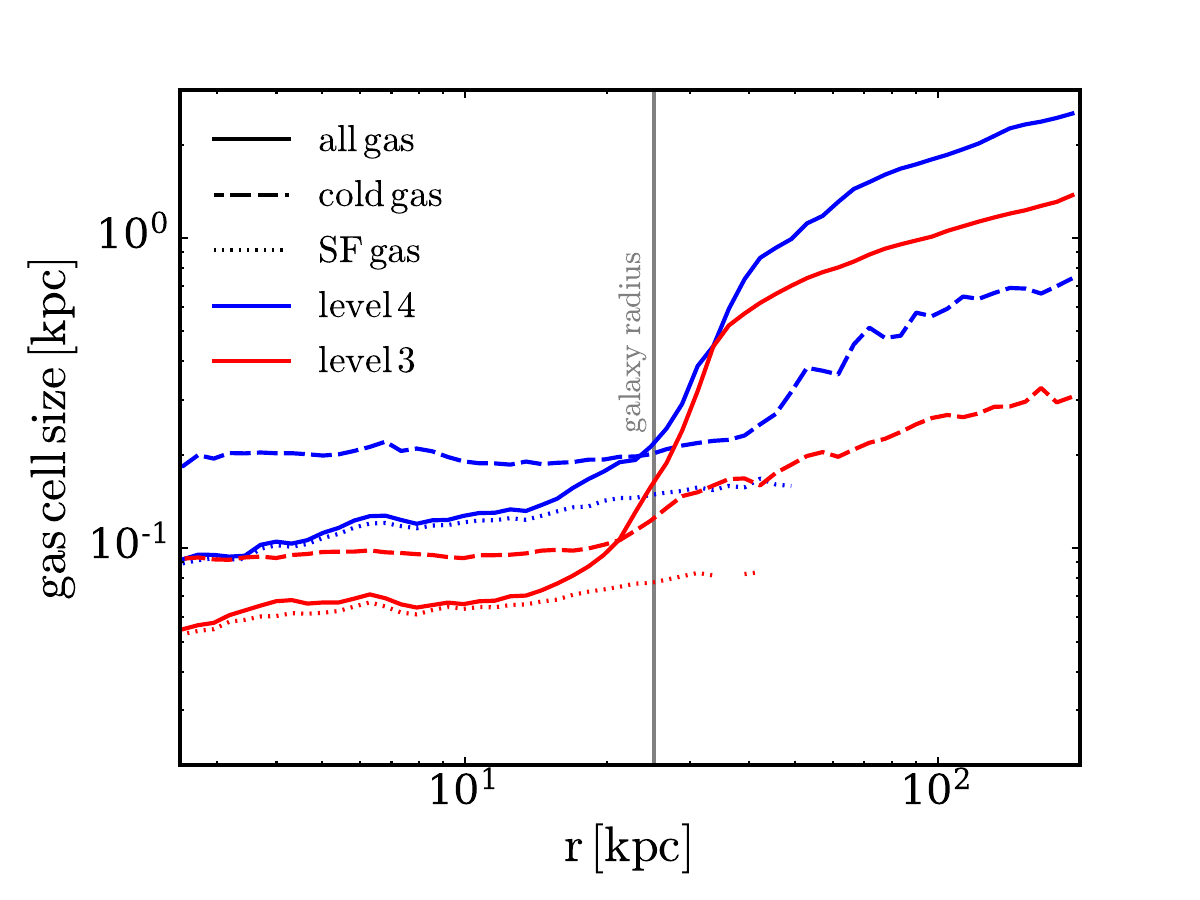}
\caption{Left: The distribution of gas cell sizes for all gas (curves) and star-forming (SF) gas (shaded) in an Auriga halo simulated at the level 4 (blue) and level 3 (red) resolutions. Right: The median size of: all gas (solid curves); cold gas, defined as gas cooler than $10^{4}$ K (dashed curves); and star forming gas (dotted curves), as a function of distance, $r$, from the halo centre. The galaxy radius, defined as 10 per cent of $R_{\rm 200}$, is marked by a vertical gray line.}
\label{gascells}
\end{figure*}

There are two different resolution ``levels'', which we often refer to as ``level 4'' and ``level 3'' which are a factor 8 (2) times different in mass (spatial) resolution\footnote{This nomenclature originates in the older `Aquarius' project \citep{Springel2008} of high-resolution dark matter-only simulations of Milky Way-sized galaxies.}. These correspond to baryonic mass resolutions of $\sim 5\times 10^4$ $\rm M_{\odot}$ and $\sim 6\times 10^3$ $\rm M_{\odot}$, respectively.  Although their respective softening lengths are approximately 375 pc (188 pc), gas cells exhibit a range of sizes. Fig.~\ref{gascells} shows the size distribution of gas cells (left panel) and the median gas cell size as a function of radius (right panel) for one of the Milky Way-mass halos simulated at level 4 and level 3 resolution levels. This figure shows that the median size of gas cells inside the galaxy is typically $\sim 150$ pc ($\sim 75$ pc) for the level 4 (level 3) resolution, and that star-forming gas cells can be as small as $\sim 30$ ($\sim 15$ pc) for level 4 (level 3). Note that, inside the galaxy radius, the median size of cold gas is generally larger than that of all gas because the latter is dominated by dense star-forming gas of higher effective temperature than that of the cold gas ($10^4$ K) at lower densities. Outside the galaxy radius, cold gas cells are, on average, up to a factor of several smaller compared to warmer gas cells because the latter is no longer dominated by star-forming gas in these regions. Table~\ref{t1} lists the numerical specifications and number of halos for each suite of simulations, and Tables~\ref{t2} to \ref{t5} list additional details for individual halos in each simulation suite.

Below, we provide details of how the halos were selected and their initial conditions generated (section~\ref{sec:ics}), the numerical methods used (section~\ref{sec:nummeth}), and the physics model employed (section~\ref{sec:model}).

\subsection{Initial conditions and halo selection}
\label{sec:ics}

The Auriga halos have been drawn from the Eagle dark matter only simulation of comoving side length 100 cMpc  \citep[L100N1504, introduced in][]{SCB15} and a dark matter particle mass resolution of $1.15\times 10^7$ $\rm M_{\odot}$. A list of the corresponding Eagle halo ID numbers for each Auriga halo are given in Appendix A of \citet{GGM17}, for reference. These halos were identified in the parent box at redshift zero using the ‘Friends of Friends’ (FOF) algorithm with the standard linking length \citep{DEF85}. We defined the centre of each FOF group as the potential minimum. For each mass range, an isolation parameter was calculated \citep[see][for details]{GGM17} and halos were randomly selected from the most isolated quartile.

The initial conditions for each of the Auriga simulations are created using the public Gaussian white noise field realization Panphasia \citep{J13}. The adopted cosmological parameters are $\Omega _m = 0.307$, $\Omega _b = 0.048$, $\Omega _{\Lambda} = 0.693$, $\sigma_8 = 0.8288$, and a Hubble constant of $H_0 = 100\, h\, \rm km \, s^{-1} \, Mpc^{-1}$, where $h = 0.6777$, taken from \citet{PC13}. Particles within a sphere of radius $4 R_{200}$ of the halo centre at redshift zero were traced back to the starting redshift $127$. This defines an amoebae-shaped region, which is sampled with a larger number of lower mass particles. Particles of higher mass were added at larger distances in order to bring down computational cost while maintaining the correct cosmological tidal field. We then split each original dark matter particle into a dark matter particle and gas cell pair, with masses set by the cosmological baryon mass fraction, and their relative separation equal to half the mean inter-particle spacing ensuring that the centre of mass and centre of mass velocity is retained. 

Finally, we remark on the contamination of the high resolution region from low-resolution dark matter particles. We define the zero-contamination zone as a sphere of radius $xR_{200}$ inside which there are no low-resolution (high-mass) dark matter particles. The vast majority of halos have values of $x\sim5\pm 1$, and all except two halos have values above 1, i.e., no high-mass particles within $R_{200}$ at redshift zero. Of these two, one has the very low value of 0.001 and is therefore omitted from the data release (see Table.~\ref{t3}), and the other has a value of 0.736 (halo 10 of the \texttt{Halos\_1e10msol/}3 suite). That concludes the description of the original Auriga initial conditions and halo selection.

It is possible to create fresh initial  conditions for the Auriga haloes using the online service {\small cosmICweb} \citep[COSMological Initial Conditions on the WEB,][]{Buehlmann2024}\footnote{ \url{https://cosmicweb.eu/documentation/api}}. CosmICweb can produce initial conditions for haloes in the Eagle dark matter only simulation volume -- including all the  Auriga haloes. These initial conditions are made using {\small MUSIC} \citep{HahnAbel2011}, which can generate initial conditions in the native file formats of many of the most widely used cosmological N-body/hydrodynamic codes.

As an illustration, \citet{Buehlmann2024} perform a dark matter only resimulation of the Auriga~6 halo.  The properties of this new version of this halo are compared to an original Auriga~6 level 4 dark matter only simulation.  There is good agreement in the final halo properties with with $M_{200}$ mass matching to better than 1\%.

\subsection{Numerical methods}
\label{sec:nummeth}

The zoom re-simulations are performed with the second-order accurate gravo-magnetohydrodynamics (MHD) code {\sc arepo} \citep{Sp10,PSB15} that we describe briefly here. {\sc arepo} is a moving-mesh code that follows MHD and collisionless dynamics in a cosmological context. Gravitational forces are calculated by a standard TreePM method \citep{Sp05}, which itself employs a fast Fourier Transform method for long range forces, and a hierarchical oct-tree algorithm \citep{BH86} for short range forces, together with adaptive time-stepping. To follow the MHD, {\sc arepo} utilizes a dynamic unstructured mesh constructed form a Voronoi tessellation of a set of mesh-generating points (the so-called Voronoi mesh), that allows for a finite-volume discretization of the MHD equations. 

{\sc arepo} generates an unstructured mesh that ensures that each cell contains a given target mass (specified to some tolerance), such that regions of high density are resolved with more cells than regions of low density. Furthermore, the mesh generating points are able to move with the fluid flow velocity, such that each cell of the newly constructed mesh moves approximately with the fluid. In this way, {\sc arepo} overcomes the Galilean non-invariance problem of standard Eulerian mesh codes and significantly reduces the advection errors that can accumulate from large relative bulk gas velocities inherent to galaxy formation \citep{Wadsley2008}. The quasi-Lagrangian characteristic of the method makes it relatable to other Lagrangian methods such as SPH, although several limitations of the SPH method are eliminated, for example, there is no artificial viscosity, and the hydrodynamics of under-dense regions are treated with higher accuracy. However, unlike SPH codes, discrete ``parcels'' of gas cannot simply be traced backwards/forwards in time by tagging a ``particle'' of a given ID at each snapshot, because conceptually gas is allowed to move between different gas cells; {\sc arepo} requires the probabilistic approach of Monte Carlo tracer particles \citep{GVN13} to trace gas flows. On the other hand, a major advantage of the finite volume scheme of {\sc arepo} is the ability to capture hydrodynamical shocks to good accuracy \citep{Schaal2015,Schaal2016}. 

\subsection{Physics model}
\label{sec:model}

The Auriga simulations are run with a physics model that includes treatments for important galaxy formation processes. The model is described in detail in \citet{GGM17}, and builds on earlier models presented in \citet{VGS13,MPS14}. Briefly, it includes: primordial and metal-line radiative cooling and heating from a spatially uniform, redshift-dependent UV background radiation field including self-shielding corrections \citep{FG09}; a pressurised model for the multi-phase interstellar medium (ISM) due to unresolved supernovae, comprising a cold, dense phase and a hot volume-filling phase \citep{SH03}; stochastic star formation in dense ISM gas above a threshold density of $n = 0.13$ $\rm cm^{-3}$ with a \citet{C03} initial mass function; stellar evolution with associated mass-loss and chemical enrichment of surrounding gas from asymptotic giant branch (AGB) stars and supernovae Ia and II; an energetic (thermal and kinetic equipartition) stellar feedback scheme powering galactic-scale gaseous outflows; seeding and growth of supermassive black holes; thermal feedback from active galactic nuclei (AGNs) in quasar and radio modes, including AGN radiation effects on nearby gas; finally, magnetic fields are seeded uniformly at the starting redshift at a comoving strength of $10^{-14}$ G (equivalent to a physical strength of $1.6\times 10^{-10}$ G) \citep{PGG17}.

\begin{table*}
\begin{center}
\begin{tabular}{ c c c c c c c c } 
 \hline
 Sim. set name/level & $M_{200}$ ($\rm M_{\odot}$) & $N_{\rm sim}$ & $M_{\rm DM}$ ($\rm M_{\odot}$) & $M_{\rm bary}$ ($\rm M_{\odot}$) & $\epsilon _{\rm comov}$ (cpc) & $\epsilon _{\rm phys} (z<1)$ (pc) & MHD/HD/DMO \\
 \hline
 \texttt{Original}/4 & $1\times 10^{12} - 2 \times 10^{12}$ & 30 & $4\times 10^5$ & $5\times 10^4$ & 750 & 375 & MHD \\ 
 \texttt{Original\_Hydro}/4 & $1\times 10^{12} - 2 \times 10^{12}$ & 4 & $4\times 10^5$ & $5\times 10^4$ & 750 & 375 & HD \\
 \texttt{Original\_DMO}/4 & $1\times 10^{12} - 2 \times 10^{12}$ & 30 & $4\times 10^5$ & - & 750 & 375 & DMO \\
 \texttt{Original}/3 & $1\times 10^{12} - 2 \times 10^{12}$ & 6 & $5\times 10^4$ & $6\times 10^3$ & 375 & 188 & MHD \\ 
 \texttt{Original\_Hydro}/3 & $1\times 10^{12} - 2 \times 10^{12}$ & 3 & $5\times 10^4$ & $6\times 10^3$ & 375 & 188 & HD \\
 \texttt{Original\_DMO}/3 & $1\times 10^{12} - 2 \times 10^{12}$ & 6 & $5\times 10^4$ & - & 375 & 188 & DMO \\
 \texttt{LowMassMws}/4 & $5\times 10^{11} - 10^{12}$ & 9 & $4\times 10^5$ & $5\times 10^4$ & 750 & 375 & MHD \\ 
 \texttt{LowMassMws\_DMO}/4 & $5\times 10^{11} - 10^{12}$ & 9 & $4\times 10^5$ & - & 750 & 375 & DMO \\
 \texttt{Halos\_1e11msol}/4 & $5\times 10^{10} - 5\times 10^{11}$ & 12 & $4\times 10^5$ & $5\times 10^4$ & 750 & 375 & MHD \\ 
 \texttt{Halos\_1e11msol}/3 & $5\times 10^{10} - 5\times 10^{11}$ & 12 & $5\times 10^4$ & $6\times 10^3$ & 375 & 188 & MHD \\
 \texttt{Halos\_1e10msol}/3 & $5\times 10^{9} - 5\times 10^{10}$ & 14 & $5\times 10^4$ & $6\times 10^3$ & 375 & 188 & MHD \\
 \hline
\end{tabular}
\end{center}
\caption{For each simulation suite: the mass range of the simulated halos ($M_{200}$); the number of simulations available ($N_{\rm sim}$); the dark matter particle mass ($M_{\rm DM}$); the baryonic particle/cell mass ($M_{\rm bary}$); the comoving softening length ($\epsilon _{\rm comov}$); and the maximum physical softening length $\epsilon _{\rm phys} (z<1)$.}
\label{t1}
\end{table*}

\subsubsection{Differences to the IllustrisTNG model}
\label{sec:illcomp}

The Auriga physics model is broadly similar to that of IllustrisTNG \citep{Pillepich2018}, with 4 key differences:

\begin{itemize}
    \item The equation of state used for the ISM: Auriga adopts the original ``stiff'' form of \citet{SH03}, whereas TNG uses a ``softer'' equation of state, which may affect the thickness of the star-forming gas layer of the disc \citep{Verma2021}.
    \item The radio mode of AGN feedback: Auriga gently adds thermal energy to random locations in the halo gas, thus inflating hot bubbles to balance X-ray losses from the halo \citep[see section 2.4 of][for more details]{GGM17}, whereas TNG uses a kinetic jet model.
    \item Stellar wind scaling: Auriga does not impose a floor for the minimum velocity of non-local stellar winds \citep[see][]{VGS13,GGM17}, whereas TNG enforces a fixed wind velocity floor of $350$ $\rm km\,s^{-1}$. This differences ensures slower winds for halos of mass $\lessapprox 10^{11}$ $\rm M_{\odot}$ in Auriga compared to TNG.
    \item Stellar yields: Auriga uses the same yield tables as the original Illustris: \citet{K10} (AGB stars); \citet{PCB98} (SNII); and \citet{TAB03,THR04} (SNIa), whereas TNG uses different yield tables for SNIa and additional yield tables for some stellar mass ranges of SNII and AGB stars \citep[see Table 2 of][for more details]{Pillepich2018}.
\end{itemize}

Although each of the differences listed above have an impact on the outcome of the simulations, the broad similarities between the models lend confidence that, for many observables, the Auriga zoom-in simulations are effectively validated on large cosmological scales/volumes.

\section{Raw Data and Data Products}
\label{dataprod}

This section contains the details of the raw data and post-processed data products that accompany this release. Information and access to current and future data is available online at \href{https://wwwmpa.mpa-garching.mpg.de/auriga/}{https://wwwmpa.mpa-garching.mpg.de/auriga}. With this paper, we release

\begin{itemize}
    \item {\bf Snapshots}: raw simulation data (gas cells, dark matter particles, stellar and wind particles, and supermassive black hole particles) at each output time;
    \item {\bf Group catalogues}: galaxy and halo properties for the same snapshot output times;
    \item {\bf Merger trees}: evolutionary tracks of dark and luminous subhalos and their properties;
    \item {\bf Initial conditions}: positions, velocities, and IDs of full matter particles at $z=127$;
    \item {\bf Accreted particle lists}: supplementary catalogues detailing additional information on the history of accreted star particles useful for Galactic archaeology studies.
\end{itemize}

In the following, we describe each data product. We note that further documentation can be found online at \href{https://wwwmpa.mpa-garching.mpg.de/auriga/dataspecs.html}{https://wwwmpa.mpa-garching.mpg.de/auriga/dataspecs}. It is important to note also that merger trees and accreted particle lists do not exist for the simulation suites  \texttt{Original\_Hydro}/3 and  \texttt{LowMassMWs\_DMO}/4. The current status of and future updates to data product availability for each simulation suite can be found online at \href{https://wwwmpa.mpa-garching.mpg.de/auriga/data_new.html}{https://wwwmpa.mpa-garching.mpg.de/auriga/data}.

\subsection{Snapshots}
\label{dataprod:snaps}

\begin{figure*}
\centering
\includegraphics[scale=0.3,trim={4cm 2cm 0 0}, clip]{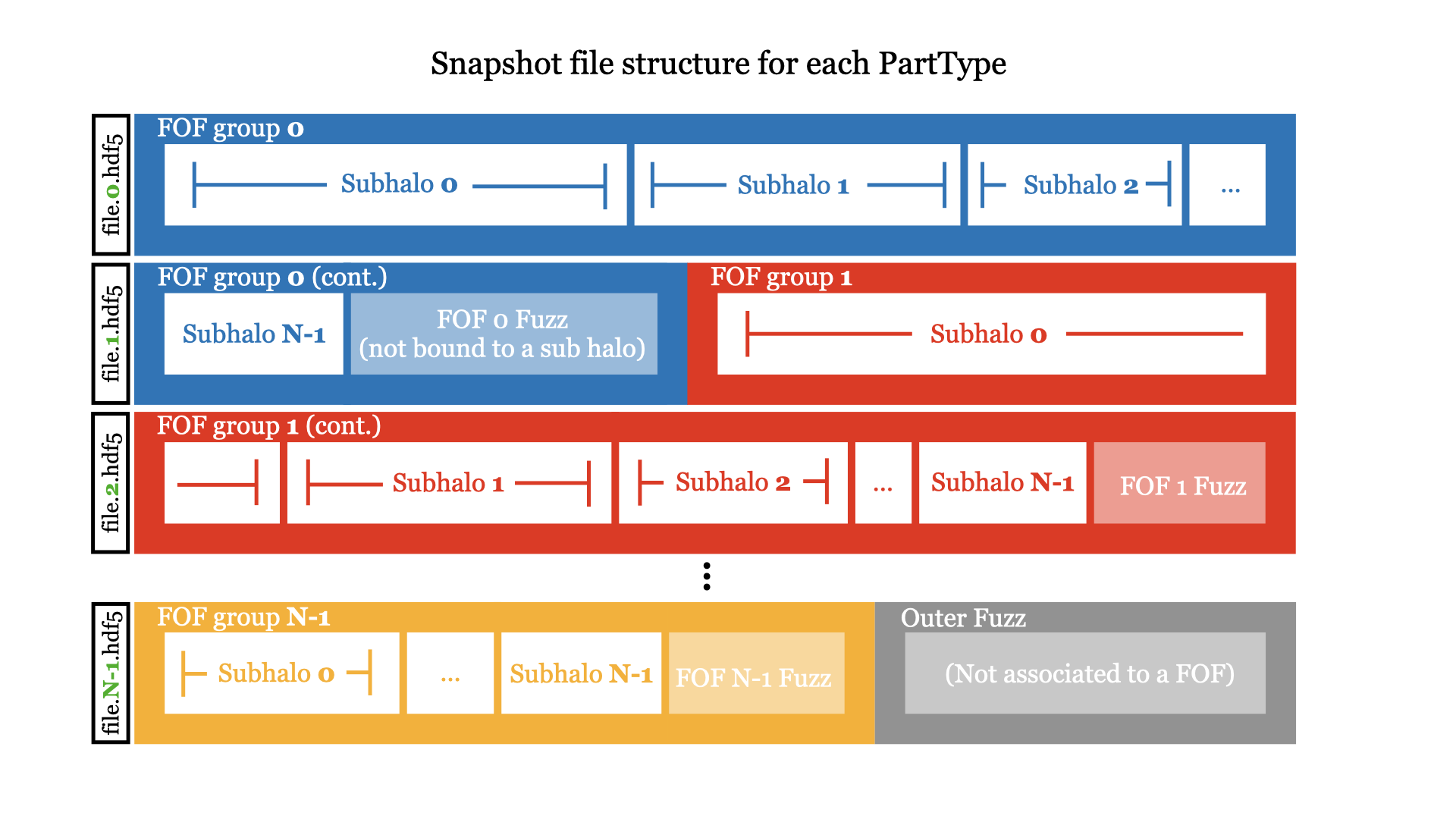}
\caption{A schematic to show how the data for a given particle type is organised in a snapshot. The order is first sorted by FOF group mass. The particles stored after the last FOF group do not belong to any group, and are said to be part of the ``fuzz''. Within each FOF group, particles are then ordered according to the SUBFIND subhalo to which they belong in order of descending subhalo mass. Similarly, within each FOF there are particles that do not belong to any subhalo according to SUBFIND, which are termed ``FOF fuzz''. Finally, within each subhalo, particles are ordered according to binding energy. Each row represents a snapshot chunk: note that particles belonging to the same FOF group or subhalo can be spread over more than one snapshot chunk. }
\label{snaps}
\end{figure*}

Each Auriga simulation stores a number of snapshots containing the raw data of each particle and cell (see Table~\ref{t1}). The particle/cell data in any given snapshot is organised according to their group/subgroup membership, as per the FoF or Subfind algorithms, and distributed across a number ($N_{\rm chunk}$) of files, or ``chunks'', denoted via \texttt{snapshot.x.hdf5}, where \texttt{x} is an integer satisfying $0 \leq \texttt{x} < N_{\rm chunk}$. For each particle type, the sort order is: GroupNumber, SubgroupNumber, BindingEnergy, where particles belonging to the group but not to any of its subgroups (``fuzz'') are included after the last subgroup. Figure~\ref{snaps} provides a schematic view of the particle organization within a snapshot, which applies to any particle type. Note that the particle distribution across chunks may mean that some FOF groups and subhalos are spread across subsequent chunks, so it is important to load all chunks for each snapshot.

\begin{table*}
\begin{center}
\begin{tabular}{ c c c c c c c c c } 
\hline
 Halo & 
 $M_{200} [\rm 10^{10} M_{\odot}]$ &
 $R_{200} [\rm kpc]$ &
 $M_{*} [\rm 10^{10} M_{\odot}]$  &
 $N_{\rm snap}$ &
 Data size (GB) &
 Comment &
 GES &
 LMC \\
 \hline
 \multicolumn{9}{c}{Level 4 resolution} \\
 \hline
 1 & 93.376 & 206.032  & 2.955 & 128 & 450 & Barred & & \checkmark \\
 2 & 191.466 & 261.757 & 9.033 & 128 & 484 & Large disc, barred & & \\
 3 & 145.777 & 239.019 & 8.617 & 128 & 327 & Large disc, unbarred & & \\
 4 & 140.885 & 236.310 & 8.045 & 128 & 435 & Unbarred & & \\
 5 & 118.553 & 223.091 & 6.920 & 128 & 273 & Barred & \checkmark & \\
 6 & 104.385 & 213.825 & 5.271 & 128 & 289 & Unbarred & & \\
 7 & 112.043 & 218.935 & 5.434 & 128 & 310 & Barred & & \checkmark \\
 8 & 108.062 & 216.314 & 3.764 & 128 & 528 & Grand design, unbarred & & \\
 9 & 104.971 & 214.224 & 6.228 & 128 & 292 & Barred & \checkmark & \\
 10 & 104.710 & 214.061 & 6.073 & 128 & 266 & Compact, barred & \checkmark & \\
 11 & 164.935 & 249.053 & 8.332 & 128 & 359 & Compact, barred, companion & & \\
 12 & 109.275 & 217.117 & 6.389 & 128 & 314 & Barred & & \checkmark \\
 13 & 118.904 & 223.325 & 6.491 & 128 & 512 & Compact, barred & & \checkmark \\
 14 & 165.721 & 249.442 & 11.118 & 128 & 691 & Barred & & \checkmark \\
 15 & 122.247 & 225.400 & 4.201 & 128 & 275 & Unbarred & \checkmark & \\
 16 & 150.332 & 241.480 & 6.838 & 128 & 629 & Large disc, unbarred & & \\
 17 & 102.835 & 212.769 & 7.913 & 128 & 545 & Compact, barred & \checkmark & \checkmark \\
 18 & 122.074 & 225.288 & 8.256 & 128 & 502 & Barred & \checkmark & \\
 19 & 120.897 & 224.568 & 5.967 & 128 & 319 & Unbarred & & \checkmark \\
 20 & 124.922 & 227.028 & 5.416 & 128 & 409 & Barred, companion & & \\
 21 & 145.090 & 238.645 & 8.232 & 128 & 565 & Unbarred & & \checkmark \\
 22 & 92.621 & 205.476 & 6.126 & 128 & 217 & Compact, barred & \checkmark & \\
 23 & 157.539 & 245.274 & 9.459 & 128 & 302 & Large disc, barred & & \checkmark \\
 24 & 149.178 & 240.856 & 7.432 & 128 & 792 & Large disc, barred & \checkmark & \\
 25 & 122.109 & 225.305 & 3.502 & 128 & 284 & Grand design, barred & & \checkmark \\
 26 & 156.384 & 244.685 & 11.218 & 128 & 741 & Misaligned disc, barred & & \\
 27 & 174.545 & 253.806 & 9.972 & 128 & 430 & Large disc, barred & \checkmark & \checkmark \\
 28 & 160.538 & 246.833 & 10.739 & 128 & 413 & Compact, barred & & \\
 29 & 154.243 & 243.553 & 10.031 & 128 & 682 & Misaligned disc, unbarred & & \\
 30 & 110.827 & 218.148 & 4.827 & 128 & 314 & Recent major merger & & \checkmark \\
 \hline
 \multicolumn{9}{c}{Level 3 resolution} \\
 \hline
 6 & 101.480 & 211.834 & 6.395 & 64 & 1004 & Barred & & \\
 16 & 150.430 & 241.530 & 9.098 & 64 & 2268 & Large disc, barred & & \\
 21 & 141.548 & 236.688 & 8.800 & 64 & 2143 & Unbarred & & \checkmark \\
 23 & 150.374 & 241.501 & 8.996 & 64 & 1044 & Large disc, barred & & \checkmark \\
 24 & 146.791 & 239.568 & 8.698 & 64 & 3042 & Large disc, unbarred & \checkmark & \\
 27 & 169.632 & 251.400 & 9.876 & 64 & 1574 & Large disc, barred & \checkmark & \checkmark \\
 \hline
\end{tabular}
\end{center}
\caption{Summary table for the \texttt{Original} Milky Way-mass simulations ($1\times 10^{12}$ ${\rm M}_{\odot}<M_{200}(z=0)<2 \times 10^{12}$ ${\rm M}_{\odot}$). The first six columns are: i) the halo name/identifier; ii) the total mass; iii) the halo radius; iv) the stellar mass; v) the number of snapshots available; vi) the simulation data volume. We also provide a brief comment on the morphology of each simulation, such as whether they are barred/unbarred \citep{Fragkoudi2024}, and specify whether they contain analogues of the {\it Gaia Enceladus-Sausage} (GES) merger \citep[as identified by][]{FBD19} and the Large Magellanic Cloud (LMC) dwarf galaxy \citep[as identified in table 1 of][]{Smith-Orlik2023}.}
\label{t2}
\end{table*}

\begin{table*}
\begin{center}
\begin{tabular}{ c c c c c c c } 
\hline
 Halo & 
 $M_{200} [\rm 10^{10} M_{\odot}]$ &
 $R_{200} [\rm kpc]$ &
 $M_{*} [\rm 10^{10} M_{\odot}]$  &
 $N_{\rm snap}$ &
 Data size (GB) &
 Comment \\
 \hline
 \multicolumn{7}{c}{Level 4 resolution} \\
 \hline
 L1 & 51.230 & 168.673 & 2.099  & 128 & 469  & Unbarred \\
 L2 & 84.380 & 199.194 & 2.704 & 128 & 700 & Barred \\
 L3 & 97.621 & 209.118 & 4.979 & 128 & 421 & Barred \\
 L5 & 67.605 & 185.005 & 3.026 & 128 & 287 & Unbarred, companion \\
 L6 & 72.729 & 189.571 & 3.885 & 128 & 763 & Unbarred \\
 L7 & 67.286 & 184.717 & 3.287 & 128 & 362 & Unbarred \\
 L8 & 84.468 & 199.266 & 5.287 & 128 & 414 & Barred \\
 L9 & 53.354 & 170.974 & 2.905 & 128 & 1100 & Barred \\
 L10 & 71.805 & 188.755 & 4.006 & 128 & 288 & Barred \\
 \hline
\end{tabular}
\end{center}
\caption{As Table~\ref{t2}, but for the \texttt{LowMassMWs} Milky Way simulations ($5\times 10^{11} \, {\rm M}_{\odot}<M_{200}(z=0)<10^{12} \, {\rm M}_{\odot}$). Note that halo L4 is heavily contaminated with low resolution particles so we omit it from the data release.}
\label{t3}
\end{table*}

\begin{table*}
\begin{center}
\begin{tabular}{ c c c c c c } 
\hline
 Run number & 
 $M_{200} [\rm 10^{10} M_{\odot}]$ &
 $R_{200} [\rm kpc]$ &
 $M_{*} [\rm 10^{10} M_{\odot}]$  &
 $N_{\rm snap}$ &
 Data size (GB) \\
 \hline
 \multicolumn{6}{c}{Level 4 resolution} \\
 \hline
 0 & 10.559 & 99.633  & 0.563 & 251 & 440  \\
 1 & 21.319 & 125.928 & 0.884 & 251 & 624   \\
 2 & 13.928 & 109.267 & 0.450 & 251 & 444   \\
 3 & 22.711 & 128.608 & 0.880 & 251 & 665   \\
 4 & 29.110 & 139.704 & 0.980 & 251 & 618   \\
 5 & 28.504 & 138.722 & 1.120 & 251 & 636   \\
 6 & 9.457 & 96.038 & 0.545 & 251 & 474   \\
 7 & 16.071 & 114.602 & 0.667 & 251 & 425   \\
 8 & 11.035 & 101.104 & 0.246 & 251 & 462   \\
 9 & 9.483 & 96.124 & 0.340 & 251 & 476   \\
 10 & 8.236 & 91.718 & 0.282 & 251 & 356   \\
 11 & 9.280 & 95.435 & 0.435 & 251 & 408   \\
 \hline
 \multicolumn{6}{c}{Level 3 resolution} \\
 \hline
 0 & 10.161 & 98.366  & 0.598 & 251 & 1420   \\
 1 & 21.091 & 125.475 & 1.176 & 251 & 3406   \\
 2 & 13.631 & 108.487 & 0.553 & 251 & 1562   \\
 3 & 23.016 & 129.183 & 1.188 & 251 & 3540   \\
 4 & 28.472 & 138.676 & 1.197 & 251 & 3326   \\
 5 & 28.169 & 138.179 & 1.381 & 251 & 3857   \\
 6 & 9.464 & 96.061 & 0.529 & 251 & 1558   \\
 7 & 15.977 & 114.386 & 0.813 & 251 & 1501   \\
 8 & 10.942 & 100.823 & 0.335 & 251 & 1611   \\
 9 & 10.197 & 98.478 & 0.405 & 251 & 2002   \\
 10 & 8.335 & 92.078 & 0.360 & 251 & 935   \\
 11 & 9.075 & 94.728 & 0.472 & 251 & 1329   \\
 \hline
\end{tabular}
\end{center}
\caption{Summary table for the 12 dwarf mass halos ($5\times 10^{10}<M_{200}(z=0)<5\times 10^{11}$ $\rm M_{\odot}$) simulations (\texttt{Halos\_1e11msol}). The first six columns are: i)~the halo name/identifier; ii) the total mass; iii) the halo radius; iv) the stellar mass; v) the number of snapshots available; vi) the simulation data volume.}
\label{t4}
\end{table*}

\begin{table*}
\begin{center}
\begin{tabular}{ c c c c c c } 
\hline
 Run number & 
 $M_{200} [\rm 10^{10} M_{\odot}]$ &
 $R_{200} [\rm kpc]$ &
 $M_{*} [\rm 10^{7} M_{\odot}]$  &
 $N_{\rm snap}$ &
 Data size (GB) \\
 \hline
 \multicolumn{6}{c}{Level 3 resolution} \\
 \hline
 0 & 1.011 & 45.582 & 2.885 & 251 & 408   \\
 1 & 0.562 & 37.479 & 0.005 & 251 & 422   \\
 2 & 4.813 & 76.679 & 234.175 & 251 & 757   \\
 3 & 0.606 & 38.429 & 0.169 & 251 & 390   \\
 4 & 0.414 & 33.850 & 0.029 & 251 & 383   \\
 5 & 0.512 & 36.343 & 0.506 & 251 & 403   \\
 6 & 2.424 & 61.001 & 30.324 & 251 & 717   \\
 7 & 0.437 & 34.459 & 0.228 & 251 & 459   \\
 8 & 1.273 & 49.224 & 30.728 & 251 & 559   \\
 9 & 3.678 & 70.103 & 71.644 & 251 & 692   \\
 10 & 0.626 & 38.850 & 0.039 & 251 & 409   \\
 11 & 2.558 & 62.116 & 22.211 & 251 & 673   \\
 12 & 0.712 & 40.549 & 3.628 & 251 & 409   \\
 13 & 0.444 & 34.640 & 0.851 & 251 & 673   \\
 \hline
\end{tabular}
\end{center}
\caption{Summary table for the 14 low-mass dwarf mass halos ($5\times 10^{9}<M_{200}(z=0)<5\times 10^{10}$ $\rm M_{\odot}$) simulations (\texttt{Halos\_1e10msol}). The first six columns are: i) the halo name/identifier; ii) the total mass; iii) the halo radius; iv) the stellar mass; v) the number of snapshots available; vi) the simulation data volume.}
\label{t5}
\end{table*}

\subsubsection{Snapshot contents}
\label{sec:snaps}

The simulation output is written in HDF5 file format in which data is organised in a clear hierarchical way. Every HDF5 snapshot contains a ``Header'' group and several particles type groups. The ``Header'' group contains information about some of the main parameters used in the simulation, such as the size of the simulation box, the cosmological parameters used, the redshift/scale factor of the snapshot, and the number of each type of particle. 

The main particle types are:

\begin{itemize}
    \item \textsc{parttype0} - gas;
    \item \textsc{parttype1} - high resolution dark matter;
    \item \textsc{parttype2} - intermediate resolution dark matter;
    \item \textsc{parttype3} - low resolution dark matter;
    \item \textsc{parttype4} - star particles and wind particles;
    \item \textsc{parttype5} - black hole particles;
    \item \textsc{parttype6} - tracer particles.
\end{itemize}

These groups contain the values for a set of fields of each type; dimensions, units, and descriptions of these are given online. The system of units is:

\begin{itemize}
    \item length: $\rm Mpc /$$h$;
    \item mass: $10^{10} \, \rm M_{\odot}$$/h$;
    \item velocity: $\rm km/s$;
\end{itemize}
where $h=0.6777$ is the Hubble parameter. Note that this system of units differs from IllustrisTNG which uses ${\rm kpc}/h$ units for length. 

Note that there are three types of dark matter particles in the Auriga zoom-in simulations, each representing particles of a different mass resolution. For the purposes of these simulations, the high resolution \textsc{parttype1} is the only relevant type of dark matter particle, as only this type of dark matter populates the Lagrangian zoom region in each snapshot. The size of this region is typically about 1-2 Mpc from the centre of the region, outside of which lie the lower resolution dark matter particles comprising the large scale Cosmic Web structure of the universe. 

\textsc{parttype6} are tracer particles, whose purpose is to link gas cells between snapshots using a Monte Carlo approach \citep{GVN13,GVZ19}. It is important to note that these particles exist only for the simulation suites \texttt{LowMassMws}/4, \texttt{Halos\_1e11msol}/4, \texttt{Halos\_1e11msol}/3, and \texttt{Halos\_1e10msol}/3. In these simulations, 1 tracer particle is assigned to each gas cell at the starting redshift.

\subsection{Group Catalogues}
\label{dataprod:cats}

Group catalogues contain higher level information about the properties of bound (sub)halos and FOF groups and the galaxies that they may host. One group catalogue exists for each snapshot, and is built via two main algorithms: Friends of Friends (FOF) and SUBFIND \citep{Springel2001}. FOF is applied to dark matter particles only which are grouped together based on a standard linking length, and other particle types are assigned to the same groups as their nearest dark matter particle. SUBFIND identifies gravitationally bound substructures within each FOF group and is run on all particle types. The objects found by SUBFIND are then bound halos that either contain luminous galaxies at their centres or are dark halos devoid of any star particle.  SUBFIND has two parameters, named \texttt{ErrTolThetaSubfind} and \texttt{DesLinkNgb}\footnote{See the Arepo code website, \hfil \href{https://arepo-code.org/wp-content/userguide/parameterfile.html}{https://arepo-code.org/wp-content/userguide/parameterfile.html}, for a description of these parameters.ml}, which we set to 0.5 and 20, respectively, for all our simulations.

Similarly to the snapshot data, this information is spread across chunks, denoted \texttt{fof\char`_subhalo\char`_tab\char`_???.x.hdf5}, where \texttt{???} represents the three digit snapshot number and \texttt{x} is again the chunk number. Each file contains a ``Header'', ``Group'' and ``Subhalo'' HDF5 group. In Auriga, the first subhalo of the first FOF group ($\rm FOF=0$; $\rm SUBFIND=0$) is usually the main Milky Way-mass halo of interest. All other subhalos (SUBFIND$>$0) in the FOF are then satellites of the main halo of that FOF group. The subhalos identified by SUBFIND are ordered in rank of descending total mass, such that the most massive subhalo of a group comes first. There are a number of FOF group and (sub)halo properties stored in the group catalogue outputs; we refer the reader to the online web page for details about these properties. 

\begin{figure*}
\centering
\includegraphics[scale=0.23,trim={0 5cm 0 0}, clip]{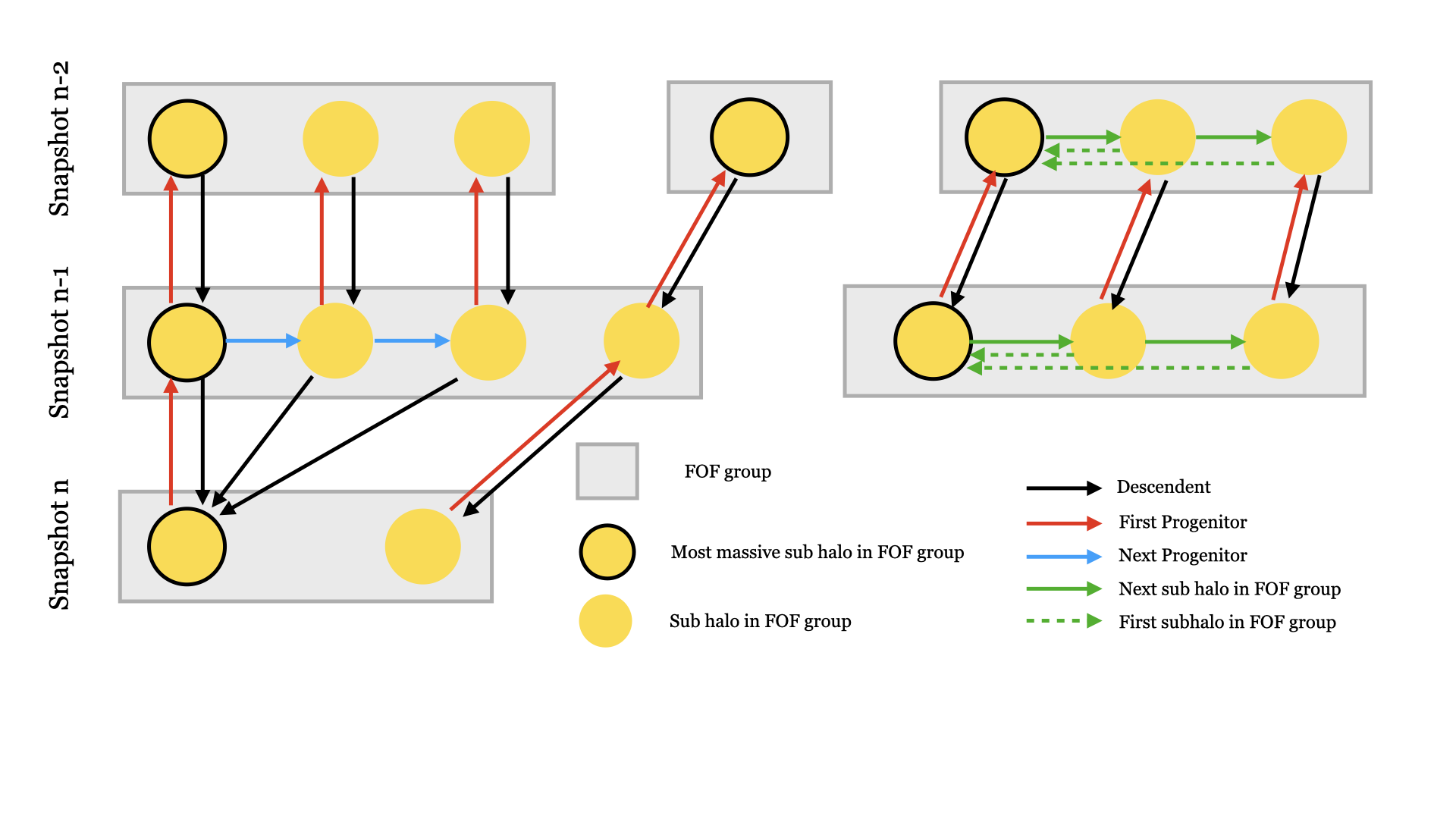}
\caption{Schematic illustrating how the linking indices given in the merger trees couple to objects in the subhalo and FOF catalgues. The most widely used linking indices are the ``FirstProgenitor'', ``Descendent'', and ``NextProgenitor''. In particular, ``Descendent'' allows one to walk forward in time and identify the same object at the next output. On the other hand, ``FirstProgenitor'' allows one to walk backward in time by identifying an object's most massive progenitor at preceding outputs. Finally, ``NextProgenitor'' moves sideways to identify other progenitors in the preceding output that have the same descendant as each other (if this is more than one, a merger has occurred). Traversing the tree in any direction (forward, backward, or sideways) can be done recursively until an index of -1 is returned, which indicates the end of the branch.}
\label{subs}
\end{figure*}

\subsection{Merger trees}
\label{dataprod:trees}

The merger trees for the Auriga simulations are produced in post-processing using the code developed by \citet{SMH05}. They provide useful linking indices that connect individual objects catalogued by FOF and SUBFIND both within the same snapshot and over different snapshots. This is important, because owing to the stochasticity in the growth of individual halos, halos of similar mass at one time may overtake each other and flip subhalo indices: for example, a satellite of the main Milky Way halo may be given the $\rm SUBFIND=0$ index if it temporarily becomes more massive. For this reason, it is best to use the merger trees to ensure the halos are consistently tracked over time.

The schematic diagram shown in Fig.~\ref{subs} illustrates how the linking indices given in the merger trees work. The most important linking indices are the ``FirstProgenitor'', ``Descendent'', and ``NextProgenitor'' ones. In particular, ``Descendent'' allows one to walk forward in time and identify the same object at the next output. On the other hand, ``FirstProgenitor'' allows one to walk backward in time by identifying an object's most massive progenitor at preceding outputs. Finally, ``NextProgenitor'' moves sideways to identify other progenitors in the preceding output that have the same descendant as each other (if this is more than one, a merger has occurred). Traversing the tree in any direction (forward, backward, or sideways) can be done recursively until an index of -1 is returned, which indicates the end of the branch. We provide an example of walking the tree on the Analysis Code \& Examples page of the data release web page.

The merger tree output is split across several HDF5 files called \texttt{trees\char`_sf1\char`_N.x.hdf5}, where \texttt{N} is the final snapshot number and \texttt{x} is the chunk number as before. In each file there are a number of groups named TreeX, where X is an integer which increases from zero to the number of tree groups in that file chunk. For the Auriga zoom-in simulations, Tree0 is the relevant tree group containing information including the ``FirstProgenitor'', ``Descendent'', and ``NextProgenitor''. We provide details of the merger tree data fields on the data release web page.

\renewcommand{\arraystretch}{1.5}

\begin{table*}
\begin{center}
\begin{tabularx}{\textwidth}{ c c c X } 
\hline
\multicolumn{4}{c}{Header} \\
\hline
 {\bf Field} & {\bf Dimensions} & {\bf Units} & {\bf Description} \\
$N_{\rm in-situ}$ & 1 & - &  Number of star particles born in-situ (here, ``in-situ'' is defined as a star particle that Subfind determines to be bound to the main galaxy at its time of birth). \\
$N_{\rm ex-situ}$ & 1 & - &  Number of star particles born ex-situ (defined as a star particle that is not bound to the main galaxy at its time of birth). \\
Time & 1 & Gyr &  The lookback time corresponding to the output file. \\
\hline
\multicolumn{4}{c}{In-situ} \\
\hline
 {\bf Field} & {\bf Dimensions} & {\bf Units} & {\bf Description} \\
\texttt{ParticleIDs} & $N_{\rm in-situ}$ & - &  The unique ID of the star particle. \\
\hline
\multicolumn{4}{c}{Ex-situ} \\
\hline
 {\bf Field} & {\bf Dimensions} & {\bf Units} & {\bf Description} \\
\texttt{AccretedFlag} & $N_{\rm ex-situ}$ & - &  Indicates whether the star particle belongs to an existing satellite (1), the main halo (0), or part of the ``fuzz'' (-9), at redshift zero. The last of these may be considered as part of the main halo at the User's discretion. \\
\texttt{BirthFofIndex} & $N_{\rm ex-situ}$ & - &  The FOF group in the corresponding snapshot in which the star particle first appeared after birth. This field can have a negative value if the star particle was not associated to any group at its time of birth. \\
\texttt{BirthSubhaloIndex} & $N_{\rm ex-situ}$ & - &  The SUBFIND subhalo index in the corresponding snapshot in which the star particle first appeared after birth. This field can have a negative value if the star particle was not associated/bound to any subhalo at its time of birth. \\
\texttt{BoundFirstTime} & $N_{\rm ex-situ}$ & Gyr & The lookback time at which the star particle first becomes bound to the main halo. This field can have a non-positive/zero value if this is not applicable. \\
\texttt{ParticleIDs} & $N_{\rm ex-situ}$ & - &  The unique ID of the star particle. \\
\texttt{PeakMassIndex} & $N_{\rm ex-situ}$ & - &  The unique tree index of the progenitor object to which the star particle belonged at the time the progenitor attained its maximum stellar mass. This field can have a negative value if this is not applicable. \\
\texttt{PeakMassInfalltime} & $N_{\rm ex-situ}$ & Gyr &  The lookback time at which the progenitor in which the star particle was born crosses inside $R_{200}$ for the first time. This field can have a negative value if this is not applicable. \\
\texttt{RootIndex} & $N_{\rm ex-situ}$ & - &  The first entry in the ``FirstProgenitor'' branch of the progenitor object in which the star particle was born. This field can have a negative value if this is not applicable. \\
\hline
\end{tabularx}
\end{center}
\caption{Additional information for star particles located within $R_{200,\rm mean}$} of the main halo at $z=0$. This information can be combined with the snapshots, halo catalogues, and merger trees to retrieve more information about accreted stars and their progenitors.
\label{tacclist}
\end{table*}

\subsection{Accreted particle lists}

We provide additional information about the star particles found within $R_{200,\rm mean}$\footnote{{$R_{200,\rm mean}$ is defined as the radius inside which the enclosed mass volume density equals 200 times the mean density of the Universe.}} of the main Milky Way-mass halo at redshift zero. This information is designed to be particularly useful for understanding the origin and accretion history of star particles that were born elsewhere and later accreted into the main halo, and that exist in the final snapshot as either disrupted debris or part of a satellite galaxy. The data fields are described in detail in Table~\ref{tacclist} and on the data release web page. Note that, for a minority of star particles, some of the ex-situ fields can have negative values if not applicable: for example, \texttt{BirthSubhaloIndex} is negative if a star particle is not associated with a subhalo at its time of birth.

\subsection{Initial conditions}
\label{dataprod:ics}

We make available the the full set of initial conditions (ICs) for level 4 and level 3 simulation suites. As described in section~\ref{sec:ics}, each set of ICs contains the coordinates, velocities, and ICs of total matter particles at $z=127$. For all DMO runs, the particles in the ICs are simply evolved forward in time, whereas for baryonic runs the IC particles are split into dark matter and gas according to the cosmic baryon fraction. Each set of ICs is spread across a number of files in binary format which can be easily read by {\sc arepo} or {\sc gadget4} \citep{Springel2021}. Instructions on how to run a set of ICs in either format using either of these codes are given on their respective websites: \href{https://arepo-code.org/wp-content/userguide/index.html}{https://arepo-code.org/wp-content/userguide/} and \href{https://wwwmpa.mpa-garching.mpg.de/gadget4/}{https://wwwmpa.mpa-garching.mpg.de/gadget4/}.

\subsection{Supplementary data}
\label{dataprod:supp}

\subsubsection{Forward modelled catalogues}

\begin{enumerate}
    \item Mock catalogues for the {\it Gaia} DR2 \citep{GHF18} for the \texttt{Original}/3 simulation suite as viewed from 4 azimuthally equidistant Solar-like positions. These include dust extinction and reddening, magnitude limits, phase-space interpolation, and error modelling. They are available to download at \href{https://wwwmpa.mpa-garching.mpg.de/auriga/gaiamock.html}{https://wwwmpa.mpa-garching.mpg.de/auriga/gaiamock} and \href{https://dataweb.cosma.dur.ac.uk:8443/gaia-mocks/}{https://dataweb.cosma.dur.ac.uk:8443/gaia-mocks/}. 
    \item Mock catalogues for the Pandas surveys \citep{Thomas2021} for the \texttt{Original}/3 simulation suite. They are available to download at \href{https://wwwmpa.mpa-garching.mpg.de/auriga/gaiamock.html}{https://wwwmpa.mpa-garching.mpg.de/auriga/gaiamock}.
    \item Synthetic UV-submm images and SEDs for 18 observer positions for each of the Milky Way-mass Auriga simulations. These are generated by the dust radiative transfer code SKIRT \citep{CampsBaes2020} and are available on the SKIRT Auriga Project website: \href{https://auriga.ugent.be/}{https://auriga.ugent.be/}.
\end{enumerate}

\subsubsection{High level data}

High level data are the outcome of analysis, such as various galaxy properties, and are often tabulated in publications. We make such high level data available in convenient ascii format that can be easily read by, for example, numpy:

\begin{enumerate}
    \item Milky Way galaxy main properties: Table 1 of \citet{GGM17} that lists properties such as the mass and scale lengths of the stellar disc and bulge of the main galaxy in the Milky Way mass simulations at redshift zero. Quantities and units are listed in the downloadable file.
    \item Milky Way galaxy HI gas disc properties: Table 1 of \citet{MGP16} that lists the mass and sizes of the HI gas discs of the main galaxy in the Milky Way mass simulations at redshift zero. Quantities and units are listed in the downloadable file.
    \item Milky Way satellite and subhalo properties: Table 1 of \citet{SGG17} that lists the mass, maximum rotation velocity, number of subhalos, and their fraction of quenched and HI-poor satellite galaxies and subhalos of Milky Way mass hosts at redshift zero. Quantities and units are listed in the downloadable file.
    \item Bar properties, including: bar formation time, bar strength, and bar length, of the Milky Way-mass simulations: Table 1 of \citet{Fragkoudi2024}. Quantities and units are listed in the downloadable file.
\end{enumerate}

In the future, more high level data will become available on the Auriga website data page.

\section{Using the data}
\label{usingdata}

\subsection{Citation policy}

We request that any article that makes use of the Auriga simulation data and/or any of its higher-level data products cite this data release paper as well as the introductory paper \citep{GGM17}. In addition, we list here some appropriate citations on particular topics using the main Milky Way set of simulations that users may find instructive: \citet{PGG17} (magnetic fields); \citet{SGG17} (satellite galaxies); \citet{MGG19} (stellar halos);  \citet{Fragkoudi+Grand+Pakmor+19,Fragkoudi2024} (stellar bars); \citet{MGP16} (HI gas properties); \citet{GWG16} (vertical structure and warps of discs).

\begin{figure}
\centering
\includegraphics[scale=0.8,trim={0 0 0 0}, clip]{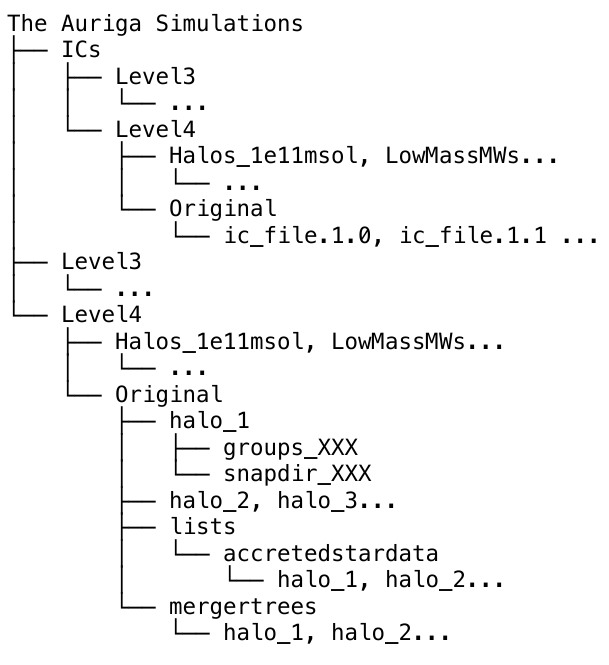}
\caption{Directory structure of the public Auriga simulation data collection, including the data products and initial conditions that we publicly release and describe in this paper.}
\label{dirstruct}
\end{figure}

\begin{figure*}
\centering
\includegraphics[scale=0.3,trim={0 4cm 0 0}, clip]{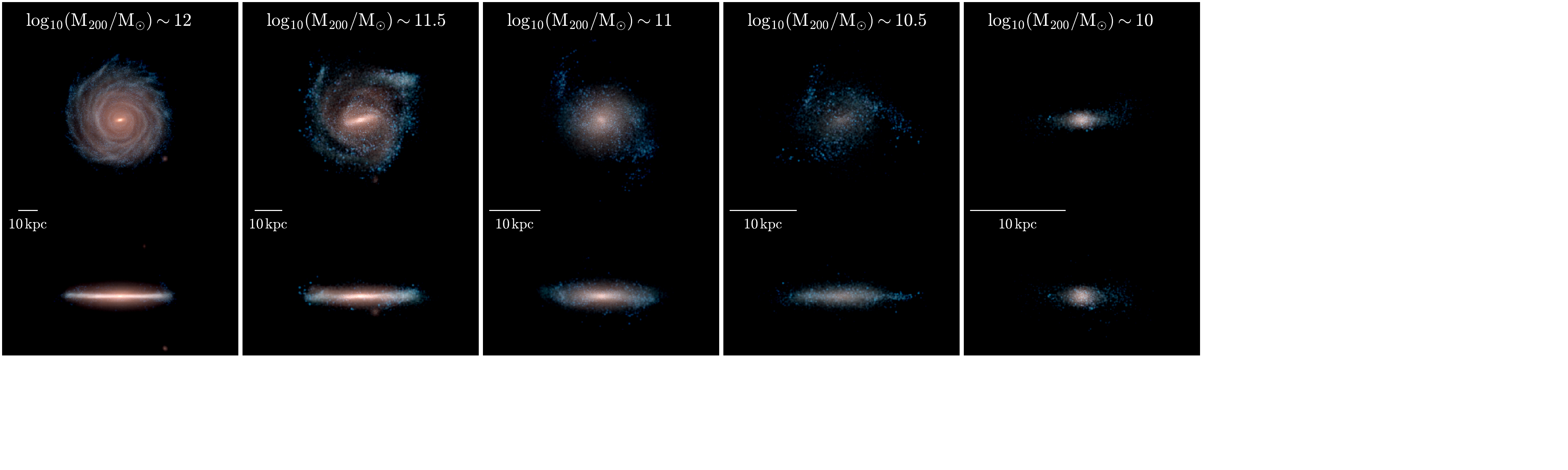}\\
\includegraphics[scale=0.3,trim={0 0 0 0}, clip]{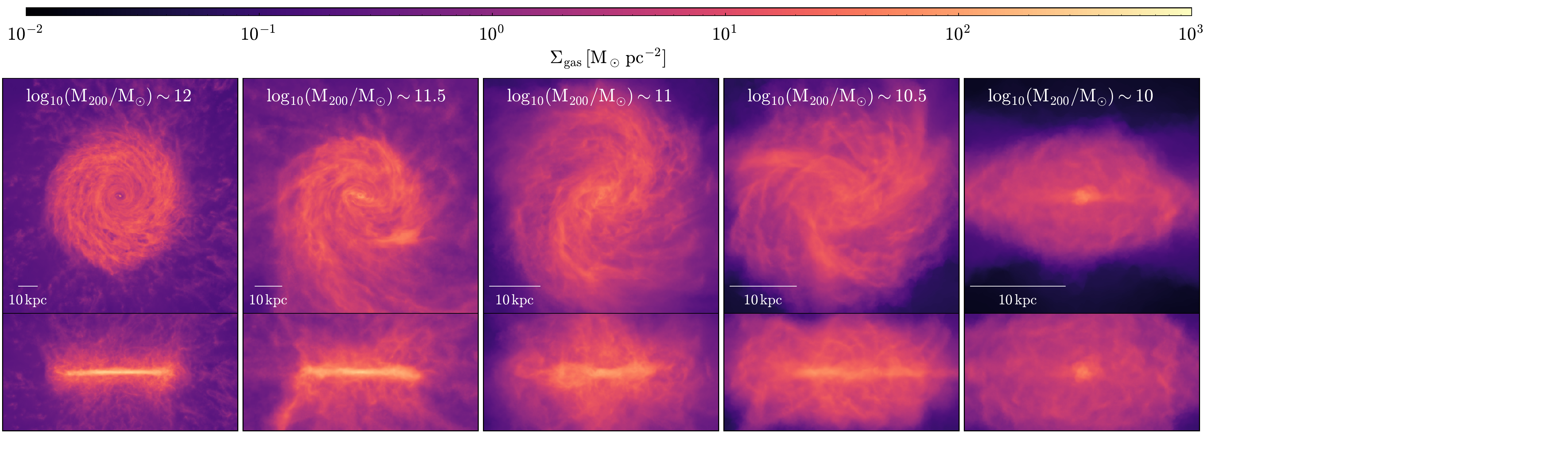}
\caption{{\it Top panels}: Face-on and edge-on stellar light projections of a selection of simulated galaxies at redshift $z=0$. The $z$-axis is aligned with the eigenvector of the moment of inertia tensor of all star particles within $0.1 R_{200}$ that is closest to the spin axis vector of those stars. We show one galaxy per 0.5 dex in halo mass to illustrate the dependence on halo mass. The scale bar is located in the bottom left corner of the face-on images in each case. {\it Bottom panels}: As for the top panels but for projections of the gas surface density.}
\label{proj}
\end{figure*}

\subsection{Data Access}

Data is accessible through the Globus file transfer service (\href{https://globus.org/}{https://globus.org/}). Globus provides a reliable and efficient system to transfer data through a user-friendly browser-based interface. The user is required to create an account and a local end-point where a data ``collection'' may be defined, which will become the location to which the data will be transferred. This can be accomplished via the browser interface after login or via the command line on machines where Globus is installed. Transfers of files and folders are then handled automatically and do not require the user to maintain connection to either end-point, i.e., transfers continue in the background and notify the user when the transfer is completed. 

The Auriga simulation data are stored in a collection named ``The Auriga Simulations'', stored at the endpoint ``MPCDF DataHub Virgotng''. A direct link to the collection can be found at \href{https://wwwmpa.mpa-garching.mpg.de/auriga/data_new.html}{https://wwwmpa.mpa-garching.mpg.de/auriga/data.html}, which will provide further instructions on how to register and access the collection through Globus. The organisation and directory structure of the released data is shown in Fig.~\ref{dirstruct}.

\subsection{Analysis tools}

We provide a user-friendly Python package for reading and analysing data which is available to download and install via a Bitbucket repository, which can be found at this link: \href{https://bitbucket.org/grandrt/auriga_public/src/master/}{https://bitbucket.org/grandrt/auriga\_public/src/master/}. To obtain this package on a local machine, one can click the ``clone'' button in the top-right of the page, and select the ``https'' option. This will provide the following command that can be copied and executed in the command line interface in the desired folder/path of the user's machine:

\begin{minted}{bash}
$ git clone https://grandrt@bitbucket.org/grandrt/
auriga_public.git
\end{minted}

These scripts require only basic Python modules available on most installations, and are intended to be simple and flexible in order to provide a starting point that the user can build on and adapt to their specific goals. In particular, they allow one to: 

\begin{itemize}
    \item Read snapshot data for a single particle type;
    \item Read the subhalo catalogues and fields;
    \item Center on the main galaxy and select particles;
    \item Rotate the galaxy to align it with the coordinate system;
    \item Convert stellar particle formation times to lookback times;
    \item Read and navigate the merger tree data to retrieve evolutionary histories of subhalos;
    \item Use the supplementary accreted star particle data to select particles from specific progenitors;
    \item Combine this additional information with the snapshot and merger tree data.
\end{itemize}

We provide some worked examples of how to perform some of these actions on the web page \href{https://wwwmpa.mpa-garching.mpg.de/auriga/analysis.html}{https://wwwmpa.mpa-garching.mpg.de/auriga/analysis.html} as well as a python notebook contained within the repository.

\subsection{Physical and numerical considerations}

\subsubsection{Numerical convergence}

Numerical convergence is a highly desirable property for simulations, because it signals that the simulation outcomes are driven by physical factors rather than numerical ones, thus increasing their reliability. Assessing convergence in galaxy formation simulations requires comparing a range of different galaxy properties of the same object at different resolution levels. 

Several published studies of the Auriga simulations have included a convergence study to understand the reliability of the simulations. In particular, \citet{GMP21} study a range of galaxy properties across 3.5 orders of magnitude in mass resolution -- spanning from a dark matter (baryonic) particle mass resolution of $m_{\rm DM}=2\times 10^7$ $\rm M_{\odot}$ ($m_{\rm b}=4.2\times 10^6$ $\rm M_{\odot}$) to $m_{\rm DM}=4600$ $\rm M_{\odot}$ ($m_{\rm b}=850$ $\rm M_{\odot}$) --  the largest resolution study to date for this type of simulation. Importantly, no parameters of the physics model (such as feedback efficiency) are varied between each simulation. A clear systematic trend was shown to be the increasing stellar mass of about $\sim 30$\% per resolution level \citep{Pillepich2019}. This kind of trend is exceedingly difficult to avoid entirely owing to higher resolution simulations resolving higher gas densities than lower resolution simulations. Nevertheless, the salient features of the star formation histories and radial density profiles are qualitatively reproduced for all resolutions above a dark matter (baryonic) particle mass resolution of $m_{\rm DM}=2.4\times 10^6$ $\rm M_{\odot}$ ($m_{\rm b}=4.4\times 10^5$ $\rm M_{\odot}$). In addition, the subhalo/satellite mass function was shown to be well-converged above this resolution. In terms of galaxy morphology, \citet{Fragkoudi2021} studied a subset of the Auriga simulations with bars; they found that bars developed independently of resolution, and that some of the dynamical properties of bars, such as the ratio of the corotation radius to bar length, $\mathcal{R}=R_{\rm CR}/R_{\rm bar}$, are within $\sim 15 \%$ for a factor 8 change in mass resolution.

An additional aspect to consider is the so-called ``butterfly effect'': the phenomenon of running the same code with the same initial conditions with a different random seed, or indeed across a different number of nodes/cpus and/or on high performance compute systems with different architectures can yield different results \citep[see, e.g.,][]{Genel2019}. Minimising these differences is a fundamental pillar holding up the predictive power of numerical simulations. Fortunately, these differences are indeed small (on the $\sim 10$\% level) for the Auriga simulations: this is shown in figure 2 of \citet{GMP21} for 7 re-simulations of the same object with different random seeds. Furthermore, the code preserves the order of operations such that two runs with the same initial conditions, code settings, and number of nodes/cores produce binary identical results, unlike several other simulation methodologies in the field. This is a helpful feature to assess the robustness of simulation predictions and results. 

\subsubsection{Usage recommendations \& known limitations}

{\it The subgrid model of the interstellar medium}: Auriga uses the \citet{SH03} two-phase model of the interstellar medium (ISM). This model  avoids explicitly modelling the highly complex range of physical processes operating in the multi-phase ISM that are challenging to model faithfully in galaxy formation simulations with current computational limitations. While the ISM model offers good convergence properties, it means that the Auriga simulations cannot be used to study the detailed structure of the multiphase ISM, such as giant molecular clouds (GMCs) and other cold clumpy structures.  

{\it Re-ionisation bug}: In the series of \texttt{Original} Milky Way-mass Auriga simulations, there is a bug in the reionisation routine. The uniform ionising UV background \citep{FG09} is suddenly switched on fully at $z=6$ instead of gradually fading in at earlier times. Therefore, the ionisation state of the IGM for $z > 6$ should be treated with caution. This has since been corrected, and does not appear to have significant effects, although a higher number of very faint galaxies are formed at early times in the corrected versions. 

{\it Black hole centering}: Black holes are seeded at the position of the halo potential minimum. With infinite resolution, drag forces imparted by dynamical friction would ensure that the black hole remained very near to this potential minimum over time. Unfortunately, this effect is not possible to resolve self-consistently in any computationally affordable simulation, which means the black hole drifts away from the potential minimum if integrated forward in time without any other outside influence. To combat this, the code regularly runs a neighbour search on gas cells and re-positions the black hole to the cell with the lowest potential. In the simulations presented in this data release, the search radius is sufficiently large and performed sufficiently frequently to ensure that black holes are generally within a softening length of the centre of the galaxy. This ensures that the site of AGN feedback remains at the centres of galaxies as intended, however, it obviously also means that our simulations should not be used to study black hole dynamics. 

{\it Tracing gas evolution}: As described in section~\ref{desc}, a consequence of the moving-mesh hydrodynamics is that the same gas is not indelibly tied to the same fluid element (cell) at different snapshots. This is because of several reasons, for example: cell velocities residual to the bulk flow mean that a fraction of gas in any given cell is allowed to be transferred across its cell face into a neighbouring cell; mesh (re-)constructions and (de-)refinements mean that cells (dis-)appear regularly. Conceptually, this means it is not possible to identify exactly the same parcel of gas at different snapshots. Instead, tracking gas in our simulations must be done with the statistical Monte Carlo tracer particle approach \citep{GVN13}. Unfortunately, tracer particles are available for only a subset of simulations (see section~\ref{sec:snaps}).

{\it Gravitational potential}: Owing to the periodic gravity in our simulations, the practise of setting the potential to zero at infinity does not apply because such a point in space does not exist. Instead, we follow the convention of setting the normalisation of the potential such that the potential is zero for vanishing density fluctuations \citep[see][for a thorough explanation]{Springel2021}. This means that one may encounter positive potential values for particles/cells at some positions. This does not affect the dynamics calculations, which involve the gradient of the potential. However, readers interested in calculating orbital energies of particles/cells with respect to the galaxy centre may wish to calculate the mean potential of a thin spherical shell located sufficiently far from the centre, e.g., at $2\, R_{200}$, and subtract it from the potential of particles/cells interior to the shell. Alternatively, one may define a zero point based on the gravitationally bound particles/cells given by SUBFIND.

\section{Comparison to observations}
\label{science}

In this section, we compare the predictions of all Auriga MHD simulations (our fiducial model) with a range of observational scalings and properties. The objectives of these comparisons serve several purposes: i) to assess the robustness and reliability of the simulations in different mass ranges and resolutions, which indicate the strengths and weaknesses of the model; ii) to make new predictions for the evolution of galaxy properties that play fundamental roles in their build-up, such as the kinematics of cool and star-forming gas; iii) to demonstrate how the simulations may be used to interpret multi-epoch observations.

\subsection{Present-day properties and relations}

\begin{figure}
\centering
\includegraphics[scale=0.45,trim={0 0 1cm 0}, clip]{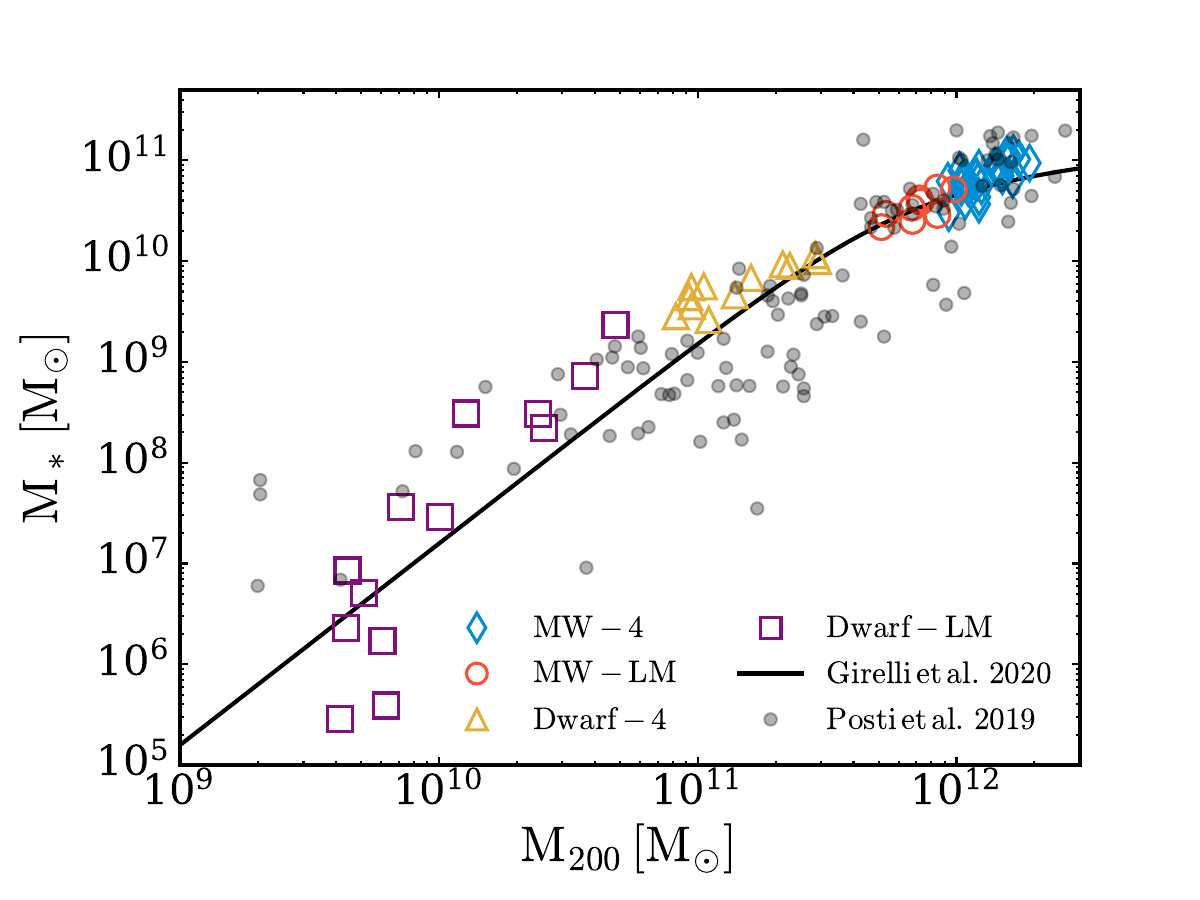}
\caption{Stellar mass versus halo mass for all of our simulated galaxies at $z=0$. For comparison, we show also the empirical relation of \citet{Girelli2020} and the values of \citet{Posti2019} derived from rotation curve fitting of nearby star-forming galaxies. }
\label{smhm}
\end{figure}

\begin{figure*}
\centering
\includegraphics[scale=0.45,trim={0 0 1cm 0}, clip]{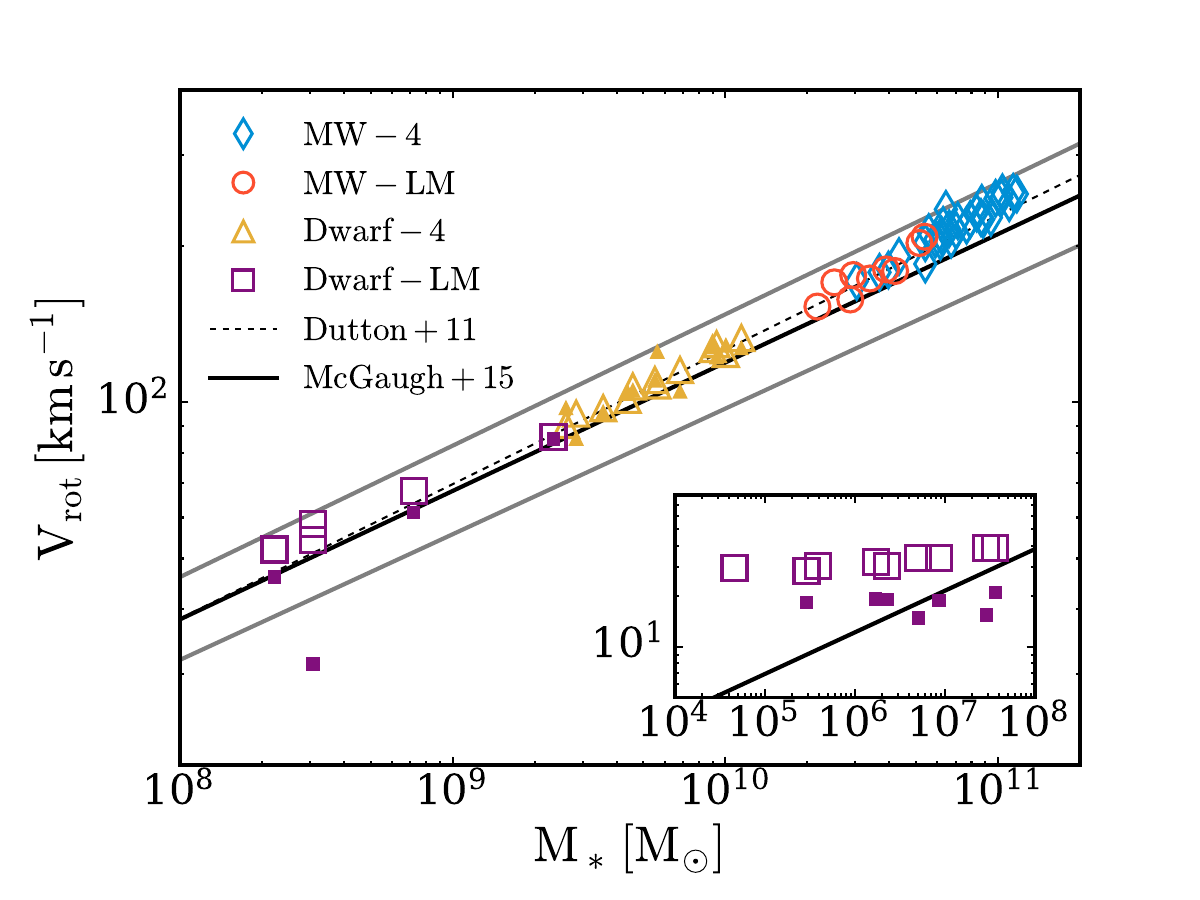}
\includegraphics[scale=0.45,trim={0 0 1cm 0}, clip]{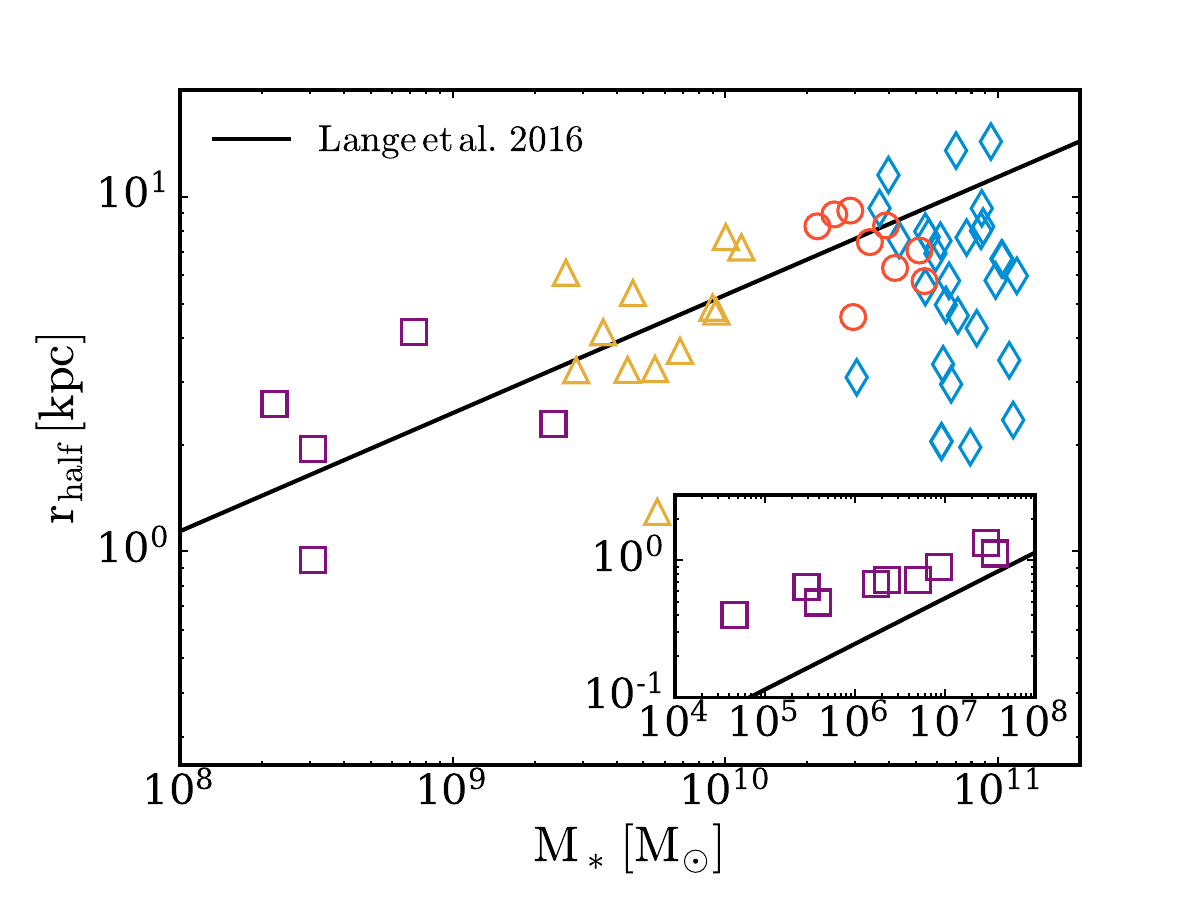}\\
\includegraphics[scale=0.45,trim={0 0 1cm 0}, clip]{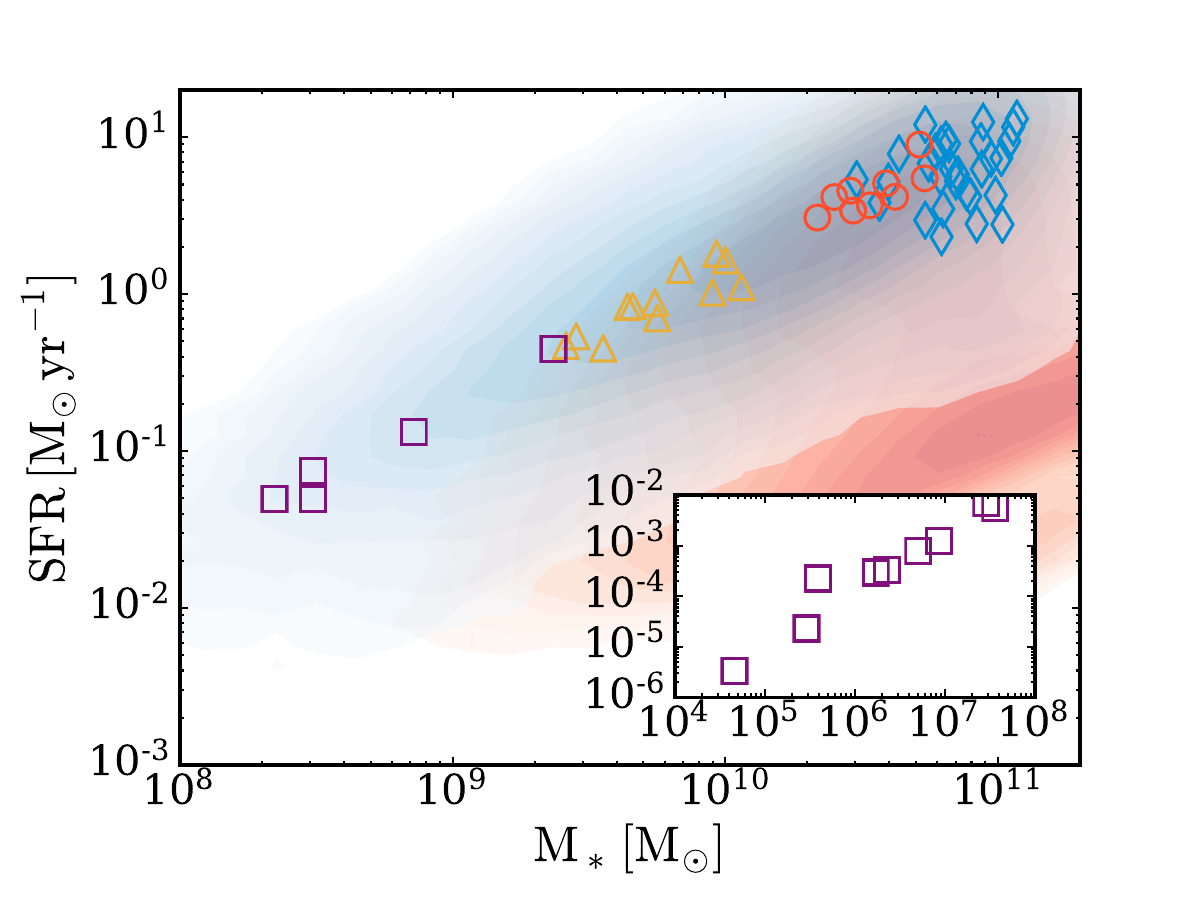}
\includegraphics[scale=0.45,trim={0 0 1cm 0}, clip]{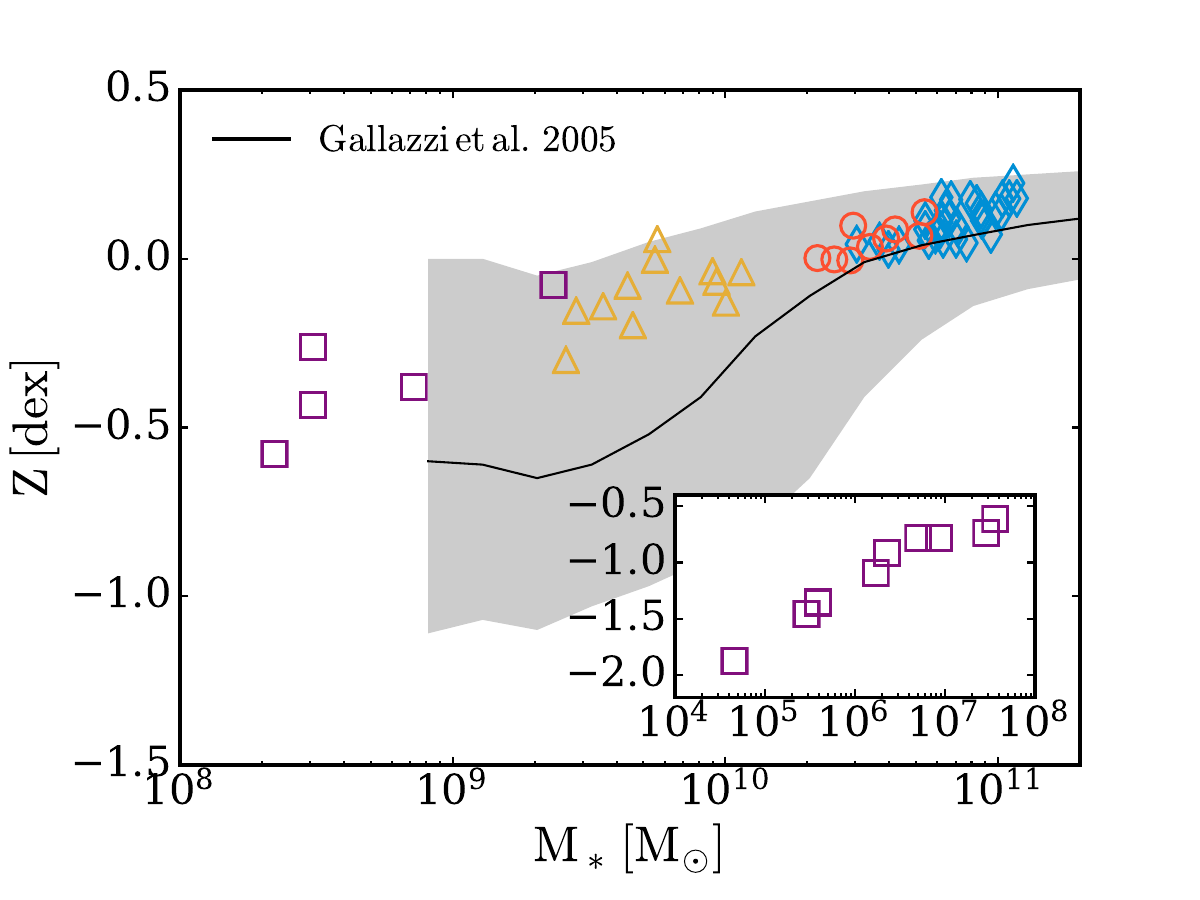}\\
\caption{Various global properties of our simulations as a function of stellar mass at $z=0$: the Tully-Fisher relation (top-left); stellar half-mass radius (top-right); star formation rate (bottom-left); and metallicity (bottom-right). In the top-left panel, $V_{\rm rot}$ for all simulations is given by the circular velocity at a radius equal to $0.1 \rm R_{200}$ (empty symbols), and for the dwarf galaxy simulations, we show an additional measure given by the mean rotation of cold gas at that radius (filled symbols). Medians (black) and spread (grey) for various observational relations are included for comparison: \citet{DBF11}; \citet{MS15}; \citet{LMD16}; \citet{GCB05}. We show also data from the SDSS MPA-JHU DR7 catalogue in the bottom-left panel.}
\label{overview}
\end{figure*}

The top row of Fig.~\ref{proj} shows face-on and edge-on stellar light projections at $z=0$; we show one halo per 0.5 dex in halo mass for brevity. The side-length of each projection is equal to $0.5 \,R_{200}$. The first and second panel show halos in the allowed Milky Way halo mass range; both clearly exhibit thin stellar discs with spiral/bar morphology. The lower mass halos in panels 3-4 show flattened, somewhat puffier discs with more irregular face-on morphology. The lowest mass halo appears more compact and spheroidal compared to the more massive galaxies, although there is a hint of a minor blue disc component. That all images display some degree of blue stellar light indicates that all galaxies are star-forming, as we quantify below. The bottom row of Fig.~\ref{proj} shows face-on and edge-on gas surface density projections at $z=0$ oriented in the same way as for the stellar light projections. The morphological trends seen in the stellar light are borne out for the gas distribution. The extent of the gas distribution is noticeably larger than that of the stars, which becomes more pronounced for lower mass halos.

In Fig.~\ref{smhm}, we show our simulated galaxies on the stellar mass-halo mass relation alongside the empirical relation of \citet{Girelli2020} as well as values given by \citet{Posti2019} that were derived from rotation curve fitting of nearby star-forming galaxies. Overall, there is good agreement between the simulations and data from \citet{Posti2019} and \citet{Girelli2020} for all mass regimes, particularly the Milky Way-mass halos. For the lowest halo masses, there is significantly higher scatter in stellar mass compared to the higher mass end, which is consistent with the findings of many other cosmological simulations of low-mass halos that often include more explicit models of the interstellar medium \citep[e.g.][]{Onorbe2015,Maccio2017,Agertz2020,Gutcke2021,Sales2022}. It is also worth to note that the scatter between the many different empirical stellar mass-halo mass relations \citep[see, for example, Figure 9 of][]{Girelli2020}, and therefore the uncertainty in the mean stellar mass for low-mass halos, is comparable to these deviations.

In Fig.~\ref{overview}, we show our simulations alongside several redshift zero observational relations that scale with stellar mass. The top-left panel of Fig.~\ref{overview} shows the Tully-Fisher relation. The rotation velocities are calculated as the circular velocity at a radius of $0.1 \rm R_{200}$, which is verified from visual inspection to be a good approximation to  the flat part of the rotation curve. Milky Way-mass halos and dwarf halos follow the scaling relation down to its lower limit of a stellar mass equal to $10^8$ $\rm M_{\odot}$. For dwarfs with stellar masses below $10^8$ $\rm M_{\odot}$, the trend flattens. This reflects the relatively large scatter in stellar mass for the halos with $M_{200}\sim 10^{10}\, {\rm M}_{\odot}$ shown in Fig.~\ref{smhm}; the rotation curves are dominated by dark matter in this regime, so large variations in stellar mass do not translate to large changes in rotation velocity compared to higher-mass halos. For the dwarf galaxy simulations, we show also a measure of $V_{\rm rot}$ given by the mean rotation of star-forming gas at the same radius ($0.1 \rm R_{200}$): this measure is systematically lower compared to the circular velocity for the lowest mass dwarfs (primarily shown in the inset) and follows the extrapolated relation. The deviation between these two measures for the low mass dwarfs is a result of the increasing importance of velocity dispersion in this regime. For higher mass galaxies, the two measures of $V_{\rm rot}$ are very similar.

The top-right panel of Fig.~\ref{overview} shows the half mass radius, $r_{\rm half}$, defined as the radius that contains half of the stellar mass. As discussed in \citet{GGM17}, the heavier Milky Way-mass halos (blue circles) lie on the observed relation but exhibit also a large scatter; some halos drop below the relation owing to effects such as mergers. The lower-mass Milky Way halos have a smaller scatter. The dwarf-mass halos continue to follow the relation down to where it ends at a stellar mass of $10^8\, \rm M_{\odot}$. If one extrapolates the observed relation, it appears that the trend in the simulations flattens for lower stellar masses. The reason for this is unlikely to be caused by a lack of spatial resolution because these low mass dwarfs were simulated at the ``level 3'' resolution, i.e., a maximum physical softening of $\sim 180$ pc (and smaller cell sizes in the star-forming gas), which is well below the half mass radii of these galaxies.

The lower-left panel of Fig.~\ref{overview} shows the present-day star formation rate of all simulations plotted alongside the blue and red sequences of the SDSS MPA-JHU DR7 catalogue. Following \citet{GGM17}, we calculate the star formation rate of stars belonging to the main halo averaged over the last $0.5\,{\rm  Gyr}$ of evolution. Two of the lowest mass dwarfs, however, formed no stars during that period, therefore we adopt the instantaneous star formation rate for these cases. All halos show a good agreement with the blue star-forming sequence. Note the absence of quenched massive galaxies in our Milky Way-mass simulations. This may be related to our AGN radio mode model and sample selection criteria (e.g. halo mass and isolation), however a detailed answer requires a dedicated study. A by-eye extrapolation of the observed sequence indicates a rough agreement even for the lowest stellar masses, which remain star-forming even if at a very low level. Finally, the lower-right panel of Fig.~\ref{overview} shows the stellar mass-metallicity relation compared with observations. Here, the Milky Way-mass halos lie relatively close to the observed median metallicity, and dwarfs lie above but within the observed scatter.

\begin{figure}
\centering
\includegraphics[scale=0.45,trim={0 0 1cm 0}, clip]{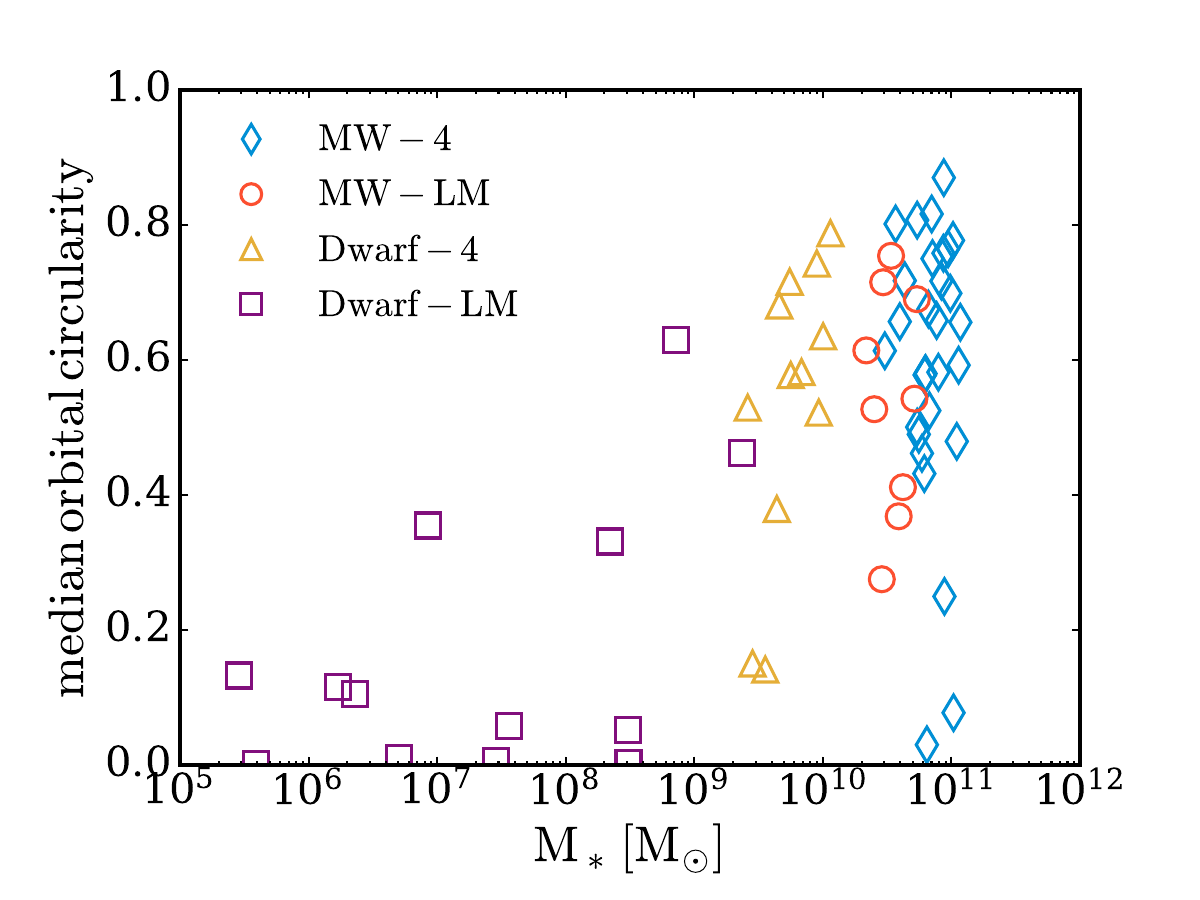}
\caption{The median orbital circularity of star particles within $0.1R_{200}$ as a function of stellar mass at $z=0$ for all simulations. }
\label{circularities}
\end{figure}

While the top panels of Fig.~\ref{overview} already show that our simulated galaxies are rotationally-supported with $r_{\rm half}$ consistent with star-forming disc galaxies, we here quantify the ``diskiness'' of our simulations using the theoretical orbital circularity quantity: values of 1 are perfectly circular orbits characteristic of galactic discs; values of 0 indicate orbits with no angular momentum more likely to be part of a spheroidal component like a spheroidal bulge or halo. We calculate the orbital circularity of star particles inside $0.1R_{200}$ following \citet{GGM17}, and take the median of these orbital circularities at $z=0$ for each simulation as a dynamical indicator of the stellar mass fraction belonging to a disc. Figure~\ref{circularities} displays the median orbital circularity of star particles as a function of stellar mass at $z=0$ for all simulations. This figure shows that there is a wide-range of values for median orbital circularity even within each simulation suite owing to a diverse range in formation histories. However, a trend of increasing circularity with increasing stellar mass is also evident, consistent with the images of Fig.~\ref{proj}.

\subsection{HI gas fractions and disc scale heights}

\begin{figure}
\centering
\includegraphics[scale=0.45,trim={0 0 1cm 0}, clip]{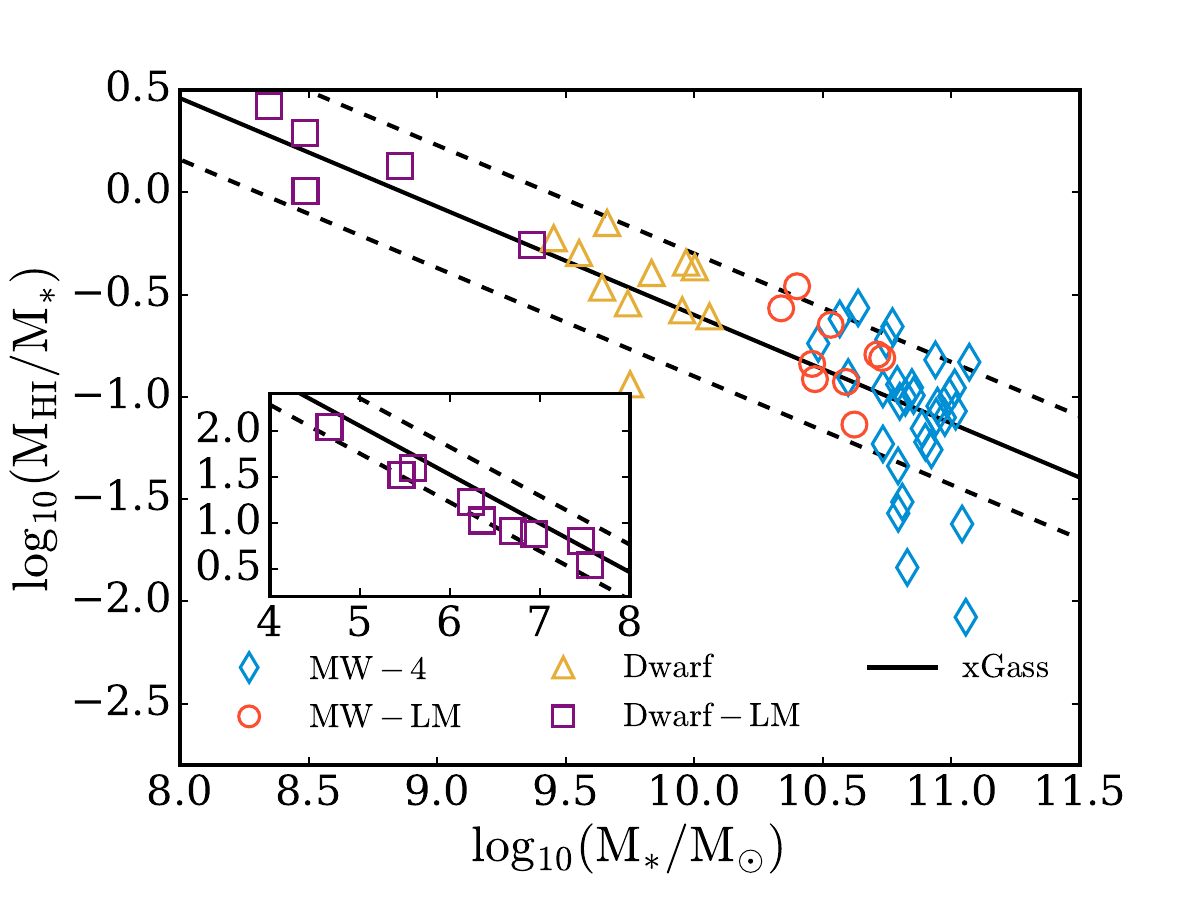}
\caption{The logarithm of the ratio between the HI mass and stellar mass for the main galaxy in each of our simulations at $z=0$. The HI main sequence relation of the star-forming xGASS \citep{Catinella2018} galaxies from \citet{Janowiecki2020} are shown by the solid black line, in addition to a 0.3 dex uncertainty shown by the dashed black lines. }
\label{MHI}
\end{figure}

\begin{figure}
\centering
\includegraphics[scale=0.45,trim={0 0 1cm 0}, clip]{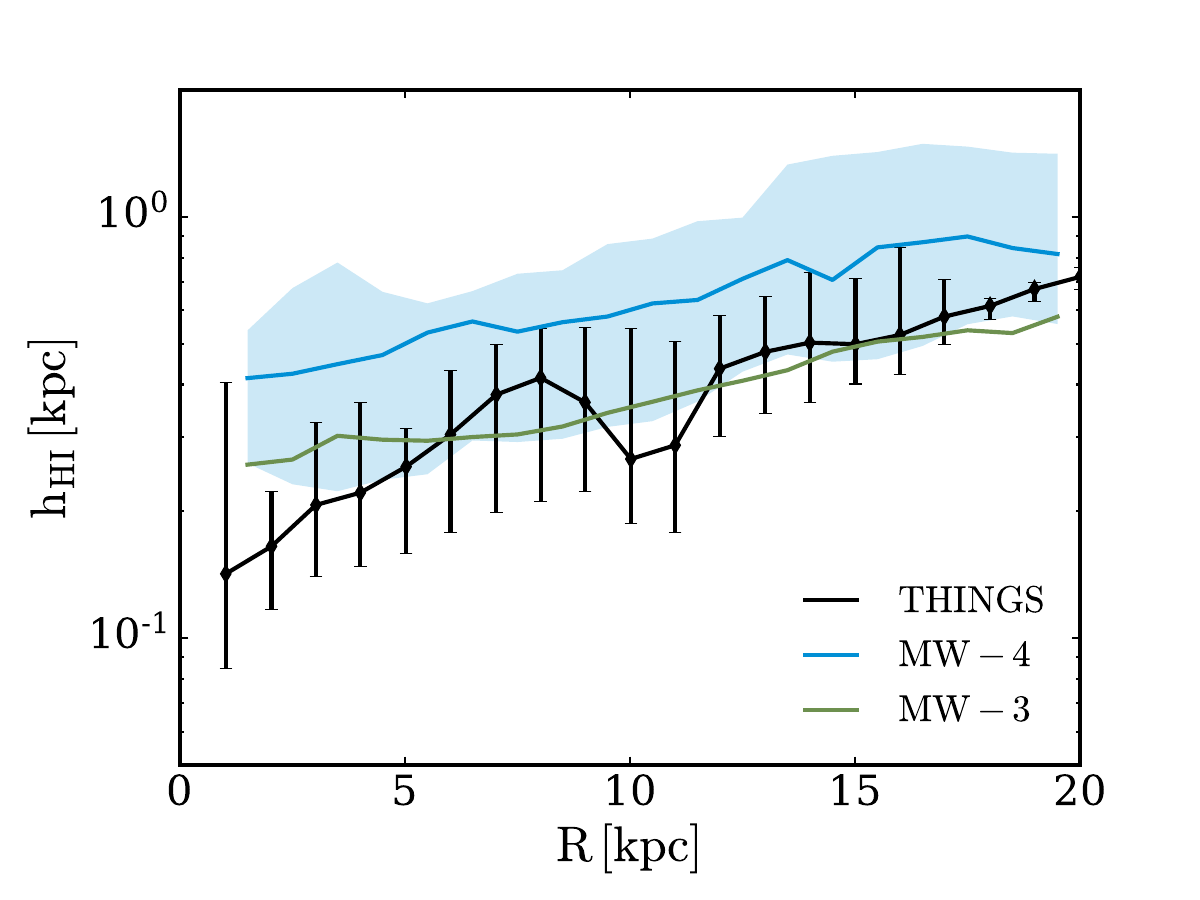}
\caption{The radial profile of the vertical scale height of HI gas at $z=0$. The blue curve and shaded regions show the median and $2\,\sigma$-equivalent percentile range for level~4 Milky Way-mass halos that exhibit undisturbed gas distributions. The green curve shows the median of the higher resolution level~3 Milky Way-mass halos. The black points and error bars are observations taken from \citet{Bacchini2019}.}
\label{HIhz}
\end{figure}

Cold gas plays a key role in the formation and evolution of star-forming galaxies. An important component of this cold gas is atomic hydrogen (HI) which acts as a reservoir of fuel for future star formation. Radio observations of HI gas in nearby galaxies are increasing in sensitivity and resolution, therefore the content and distribution of galactic HI are key observables for theoretical models of galaxy formation to reproduce. In this section, we analyse the HI gas content and distribution of our simulations in the context of prior work on this topic.

We begin by calculating the HI mass of the $i$-th gas cell by first calculating the molecular hydrogen gas fraction, $f_{{\rm mol},i}$, using an empirical relation based on the ratio of molecular to atomic hydrogen column density \citep{Blitz2006} with two parameters chosen by \citet{Leroy2008}. We then calculate the neutral gas fraction, $f_{{\rm neut},i}$, directly from the simulation output. With this information, we finally compute the HI mass as: $M_{{\rm HI},i} = (1-f_{{\rm mol},i}) f_{{\rm neut},i} X_i M_i$, where $X_i$ is the hydrogen mass fraction and $M_i$ is the total gas mass of the cell. The full procedure is described in detail by \citet{MGP16}.

In Fig.~\ref{MHI}, we show the logarithm of the ratio between the HI mass and stellar mass for the main galaxy in each of our simulations at $z=0$. Here, $M_{\rm HI} = \Sigma _{i=0}^{N_{\rm gal}} M_{{\rm HI},i}$, where $N_{\rm gal}$ is the number of gas cells inside $0.1 R_{200}$. We see that the simulations track the observed HI main sequence relation of \citet{Janowiecki2020} across the whole stellar mass range. Apart from a few Milky Way-mass galaxies that are found at relatively low HI mass fractions, the simulations generally lie comfortably within the 0.3 dex uncertainty. This agreement for the Milky Way-mass halos is consistent with our previous analysis with earlier datasets \citep{MGP16}.

In terms of the distribution of HI gas, \citet{MGP16} already showed that our Milky Way mass halos reproduce the observed mass-diameter relation well. Here, we extend the analysis of the HI distribution by calculating the vertical scale height of HI gas, $h_{\rm HI}$, for a series of radial annuli up to about 20 kpc from the centre, at $z=0$. To calculate $h_{\rm HI}$ for a given radial annulus, we bin gas cells within 5 kpc of the galactic midplane in terms of their absolute height, using bin sizes of about 250 pc (comfortably resolving the typical gas cell sizes in the galactic discs; see Fig.~\ref{gascells}), and calculate the HI vertical density profile. We then fit a $\rm sech^2$ curve with a normalisation and scale height parameter to the profiles, and accept only values with chi-squared fractional errors of less than 10\%. 

We apply the procedure described in the previous paragraph to a subset of the level 4 Milky Way sample after pruning out galaxies that are clearly visually disturbed. The median and 1-sigma scatter of $h_{\rm HI}$ as a function of radius for these simulations are shown in Fig.~\ref{HIhz}, alongside observations from THINGS \citep{Bacchini2019}. We see that the median $h_{\rm HI}$ of the simulations lies along the upper edge of the observational range, and that the scatter of both simulations and observations overlap. However, the radial profile of the median scale height for the 6 high-resolution (level 3) Milky Way simulations show lower values that lie close to the the median observational trend. Although the level 3 simulations are only a subset of the level 4 suite, Fig.~\ref{HIhz} indicates that better resolved gas near the midplane helps to produce the lower HI scale heights, which incidentally confounds the constraining power of such measurements on the physics of galaxy formation models.

\subsection{Kinematic evolution of star-forming gas}

The recent surge in advancement in high-redshift observations (e.g.~JWST, KMOS, DEEP2) and Galactic surveys (e.g.~Gaia, APOGEE) is enabling astronomers a closer look at how galactic discs formed, beginning potentially from the earliest epochs. With the availability of rotation velocities and velocity dispersions of H$\alpha$ measurements of galactic gas at different redshifts, simulations may now help connect these epochs across cosmic time to understand when discs were first established, and whether or not they formed ``upside-down''. In this section, we quantify the level of ``diskiness'' in our simulations and compare the kinematic evolution of the star-forming gas alongside some observations.

\begin{figure}
\centering
\includegraphics[scale=0.45,trim={0 0 1.8cm 0}, clip]{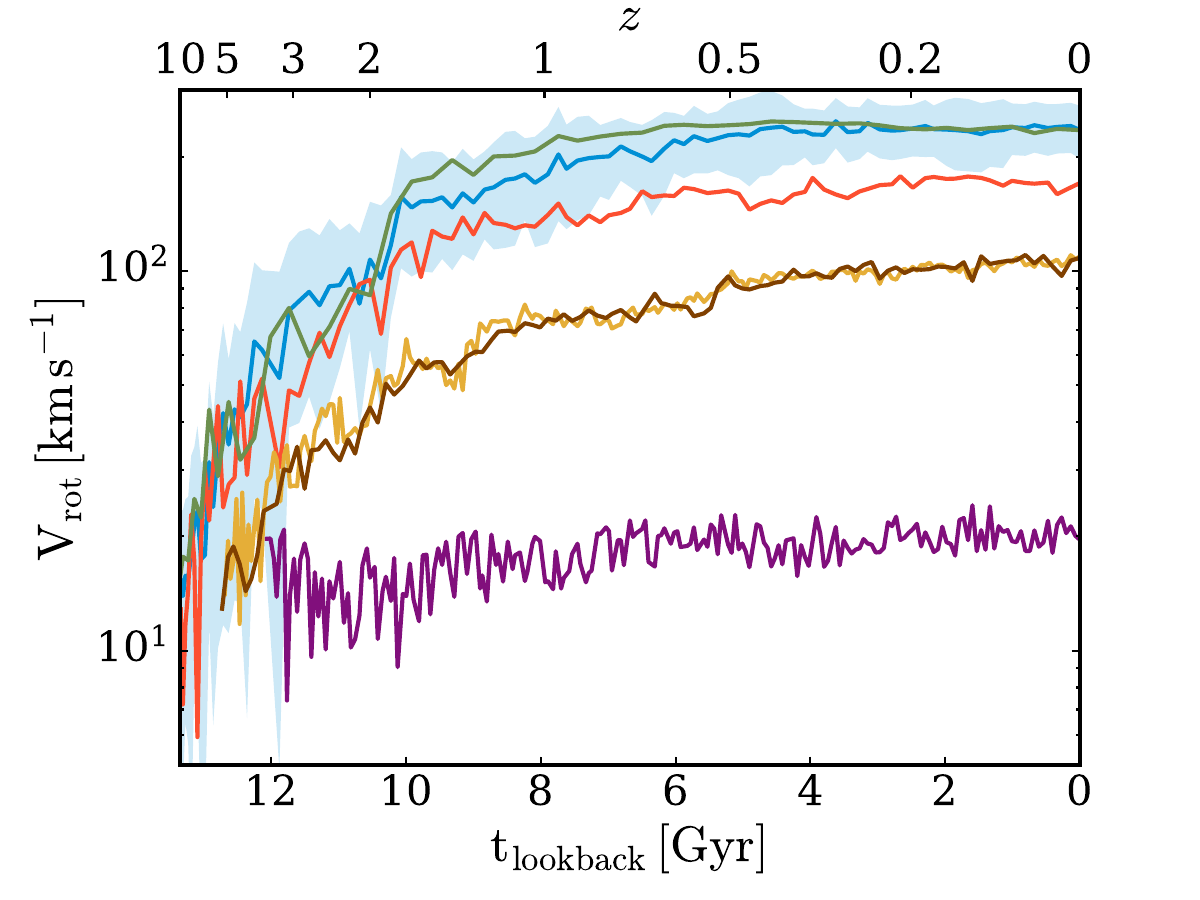}\\
\includegraphics[scale=0.45,trim={0 0 1.8cm 0}, clip]{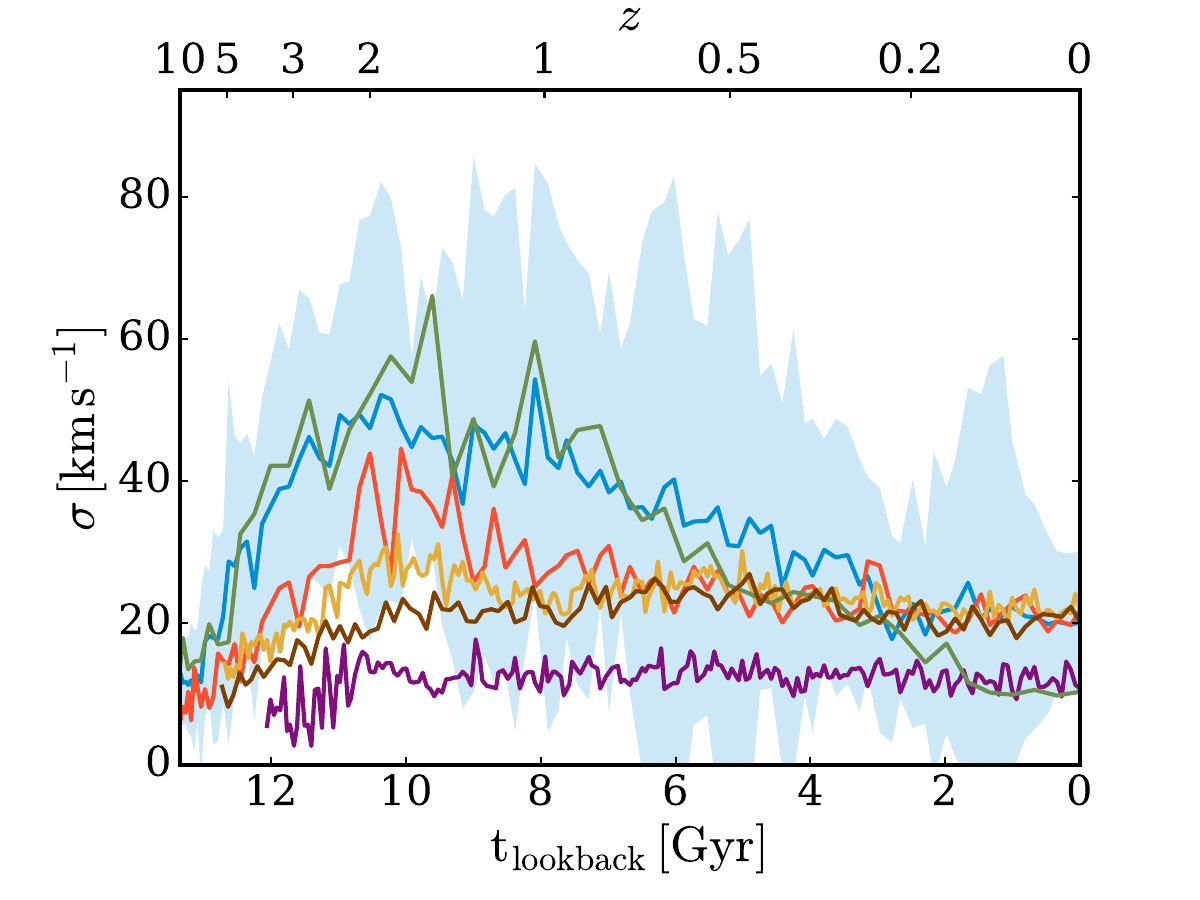}\\
\includegraphics[scale=0.45,trim={0 0 1.8cm 0}, clip]{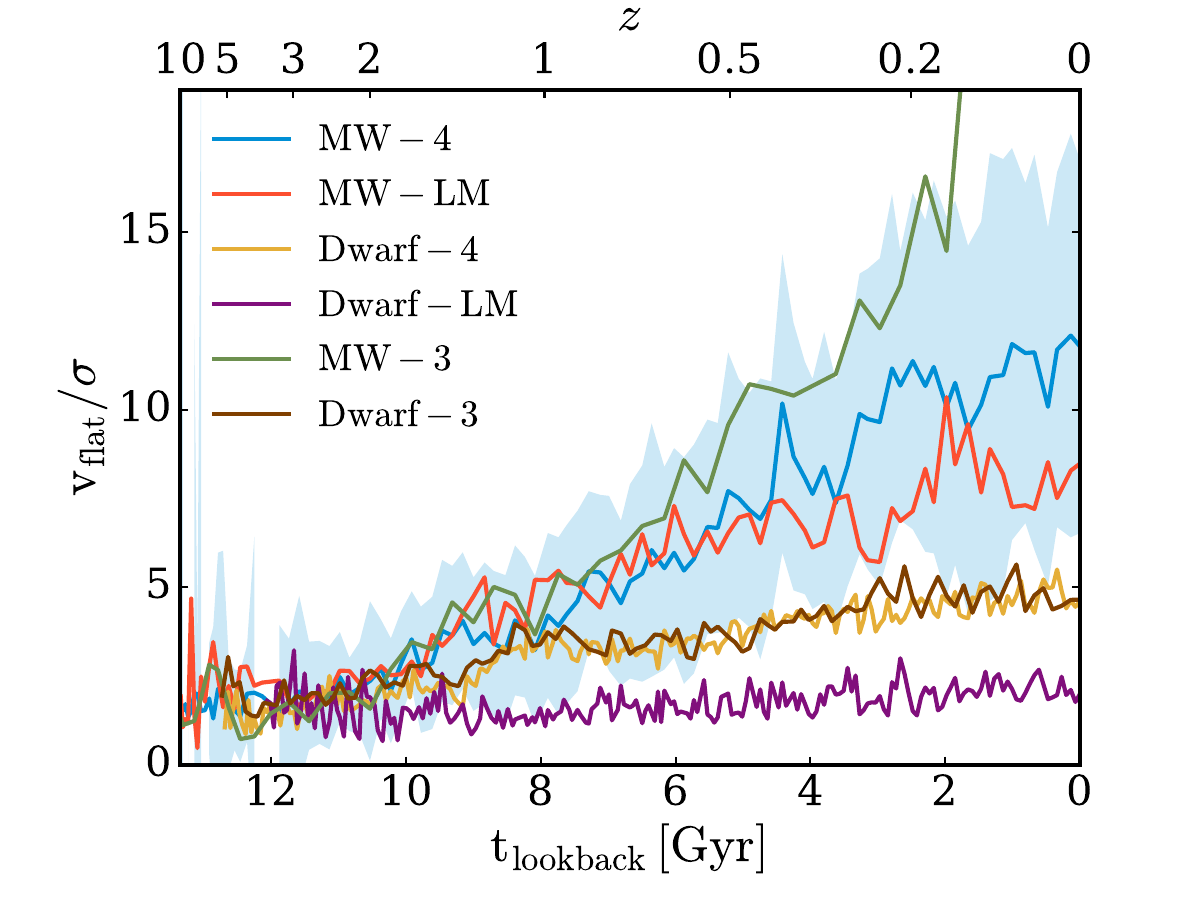}
\caption{Evolution of the rotation velocity (top panel), vertical velocity dispersion (middle panel), and the ratio of these quantities (bottom panel) of the star-forming gas for all simulations. The median for each simulation group is shown by the solid curves, and the 1-$\sigma$ scatter of the larger \texttt{Original}/4 suite by the shaded regions.}
\label{gasev}
\end{figure}

\begin{figure}
\centering
\includegraphics[scale=0.45,trim={0 0 1.5cm 0}, clip]{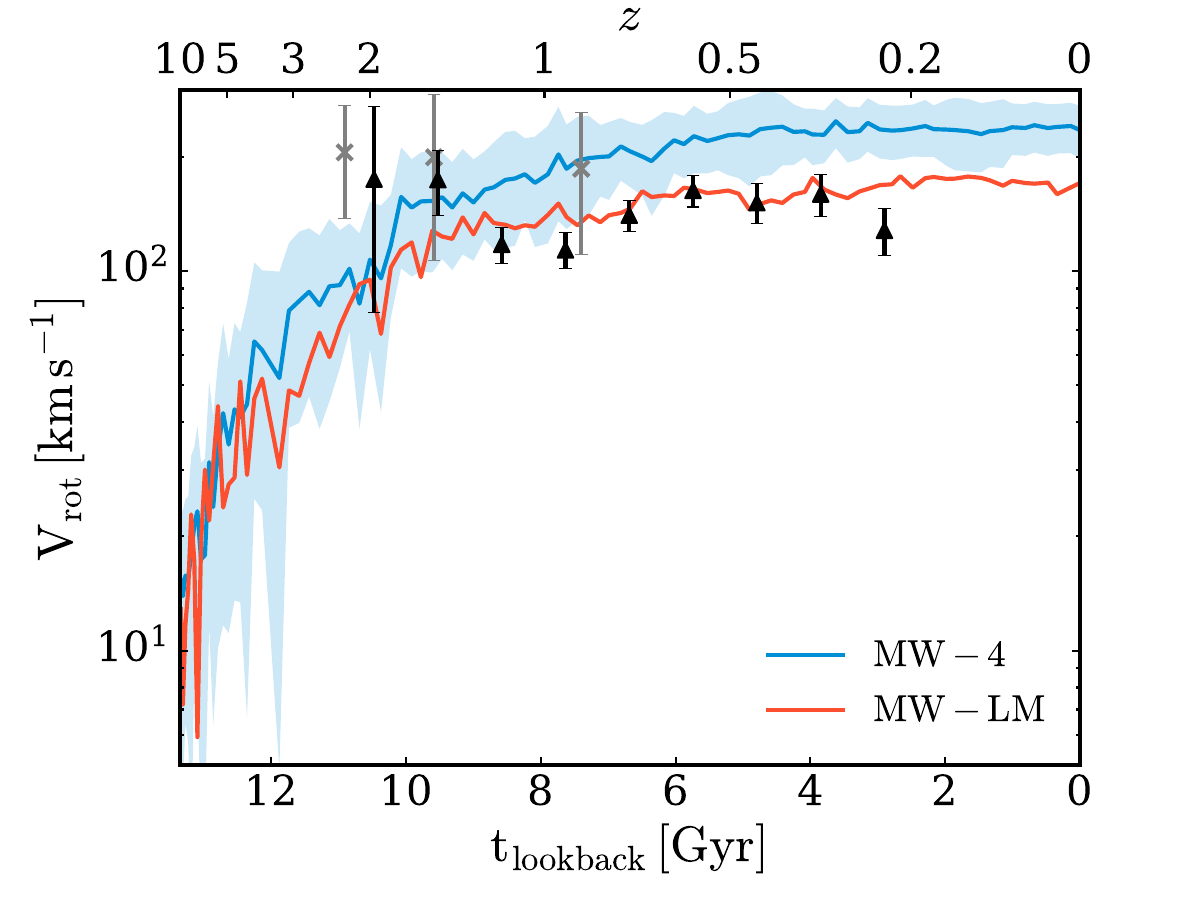}\\
\includegraphics[scale=0.45,trim={0 0 1.5cm 0}, clip]{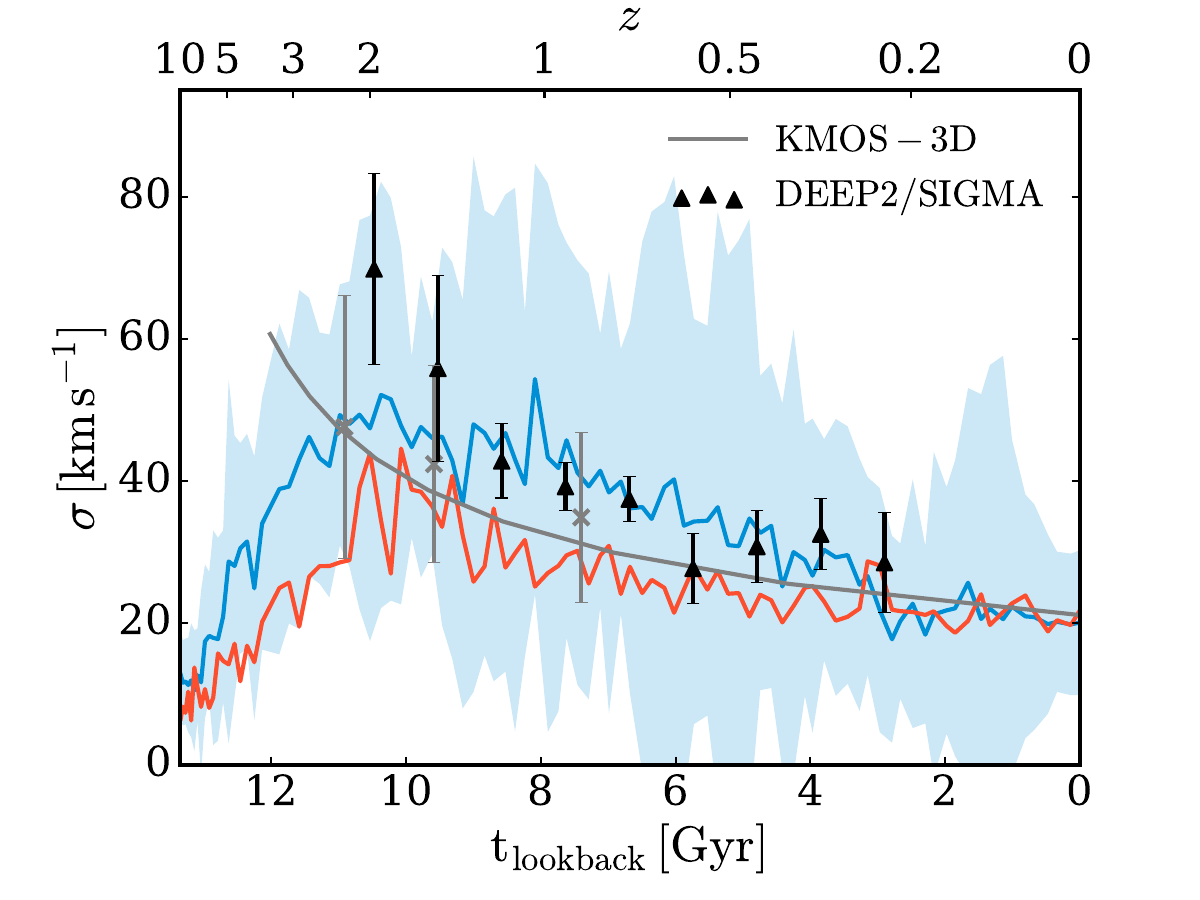}
\caption{Evolution of the rotation velocity (top panel) and vertical velocity dispersion (bottom panel) of the star-forming gas for the Milky Way-mass simulations (as shown in Fig.~\ref{gasev}). We show observational data points from \citet{Simons2017} for galaxies in the stellar mass range $10^{10}$-$10^{11}$ $\rm M_{\odot}$, and the binned KMOS-3D data (points and error bars indicate the mean and standard deviation) as well as the best-fitting relation from \citet{Uebler2019}.}
\label{gasevobs}
\end{figure}

The build up of star-forming gas discs is a major driver in shaping stellar discs at present-day. Two important quantities used to diagnose and measure the presence of gas discs are the rotation velocity and velocity dispersion of star-forming gas: the former measures ordered rotation and the latter random motion, therefore the ratio between these quantities can be regarded as a measure of rotational/pressure support a system has. We calculate these quantities at each snapshot by first aligning and rotating the stellar disc into our coordinate system (as for the projections in Fig.~\ref{proj}), then measuring the mean rotation and vertical velocity dispersion of gas in radial annuli between 1 and 2 times the stellar half-mass radius. The rotational velocity is then taken as the maximum velocity in this range whereas the velocity dispersion is averaged over the annuli. This procedure is similar to what is done in \citet{Pillepich2019}. 

In Fig.~\ref{gasev}, we show the evolution of the rotation velocity (top panel), velocity dispersion (middle panel), and the ratio between these two quantities (bottom panel) of the star-forming gas across our simulation suites. For Milky Way-mass halos, we see that the rotation velocity begins building up from very early times and, on average, reaches rotation velocities of $\sim 100$ $\rm km\,s^{-1}$ at around $z\sim 2$ (when the stellar mass is, on average, about $10^{10} \, \rm M_{\odot}$), and continues to increase until reaching steady values of just about $230 \, \rm km\,s^{-1}$ for the heavier Milky Way-mass halos ($1 < M_{200}/[10^{12}\,\rm M_{\odot}] < 2$) and about $170 \, \rm km\,s^{-1}$ for the lighter Milky Way-mass halos ($5 < M_{200}/[10^{11}\,\rm M_{\odot}] < 10$). For redshifts $z\sim 10$ to $2$, the virial temperature of the halo is sufficiently low to allow for direct cold-mode accretion, which drives a turbulent velocity field in the halo centre (incidentally, this creates also a turbulent dynamo that exponentially amplifies the magnetic field, see \citealt{PGG17} for a detailed study). This acts to increase the velocity dispersion (middle panel) until the halo grows sufficiently massive to establish the hot halo and cold flows around the virial radius are shock-heated. Gas accretion subsequently transitions to a more gradual hot-mode phase and enters a period of ``disc settling'' characterised by gradually decreasing velocity dispersion for $z\lesssim 2$ on average. This evolution translates into $v/\sigma$ ratios (bottom panel) that steadily increase from average values of about 2 prior to the establishment of the hot halo at $z\sim 2- 3$ to present-day values larger than 10 for the heavier Milky Way-mass halos and about 8 for the lighter Milky Way-mass halos; this is partly because several of the lighter Milky Way-mass halos experience significant mergers at late times which heat the disc. Indeed, part of the scatter in the heavier Milky Way-mass halos is caused by the same phenomenon. 

Figure~\ref{gasev} shows the evolution of the level 3 Milky Way-mass halos also. It reveals that they undergo a kinematic evolution very similar to that of the level 4 set for epochs earlier than $z\sim 2$. Subsequently, the level 3 simulation set differs from the level 4 simulation set in the following ways: the rotation velocity of the former reaches its maximum value at around $z\sim 0.6$ compared to $z\sim 0.4$ of the latter; during the same epoch, the median velocity dispersion of the former drops to values of approximately $10\, \rm km \, s^{-1}$ (and can be as low as $5 \, \rm km\,s^{-1}$ for individual galaxies) at $z=0$ compared to $20\, \rm km \, s^{-1}$ for the latter. These differences culminate in much larger $v/\sigma$ ratios that  rise steeply during this epoch to values of $v/\sigma \gtrsim 20$ at $z=0$. Similarly to the HI disc scale heights discussed in the previous section, these deviations are likely the result of higher numerical resolution of the level 3 set, and because this is a subset of the level 4 simulations that have not experienced late-time significant merger events and tend to have particularly dominant discs.

With respect to the high-mass dwarf galaxy simulations, we see smooth increase in rotation velocity up until values of $100 \, \rm km\,s^{-1}$ attained at $z=0$ -- about 10 Gyr later compared to the Milky Way-mass simulations. The velocity dispersion increases from early times until $z\sim 2$, at which time it remains approximately constant at just above $20 \, \rm km\,s^{-1}$. As such, $v/\sigma$ gradually increases until reaching a maximum (on average) of about 5 at $z\sim 0$. The evolution of all these quantities is quantitatively very similar for both level 4 and level 3 versions of these high-mass dwarfs, which indicates good numerical convergence. The low-mass dwarf simulations are characterised by rotation velocities and dispersions of about $15 \, \rm km\,s^{-1}$ and $10 \, \rm km\,s^{-1}$ on average, which are established shortly after their formation and evolve extremely weakly with time. 

In Fig.~\ref{gasevobs}, we plot rotational velocities and dispersions of galaxies derived by H$\alpha$ observations of galaxies with stellar mass range of $10^{10}$-$10^{11}$ $\rm M_{\odot}$ from the DEEP2 and SIGMA survey \citep{Simons2017}, in addition to binned KMOS-3D data and fit \citep{Wisnioski2019,Uebler2019}, alongside the values from the Milky Way-mass simulations. Even though these data points do not represent the continuous evolution of a group of galaxies in the same way as our simulations, the relatively broad stellar mass range of the observations mean that we may expect a rough equivalence for significant periods of evolution. Therefore, we include these observations to guide the eye for a rough comparison. The star-forming gas in our Milky Way-mass galaxies seem to build rotation and ``settle'' in (at least qualitative) accordance with what is suggested by these observations. Exciting new measurements of gas kinematics at very high redshift ($z\gtrsim6$) using JWST/JADES are just now becoming available: \citet{deGraaff2023} presents 6 data points with velocity dispersions ranging between $\sim 20$ and $\sim 70$ $\rm km\, s^{-1}$. Although this is a small number of data points, these measurements hint at very interesting future statistical comparisons between observations and simulations at the earliest epochs of galaxy formation.

\section{Summary}
\label{summary}

We have publicly released data from the Auriga project, including raw snapshots, group catalogues (FOF halos and SUBFIND subhalos), merger trees and post-processed high-level data products. The data are available to browse and download via the Auriga website: \href{https://wwwmpa.mpa-garching.mpg.de/auriga/data_new.html}{https://wwwmpa.mpa-garching.mpg.de/auriga/data}, which also contains detailed descriptions of the data. We provide some basic Python-based analysis scripts housed in a public bitbucket repository available at \href{https://bitbucket.org/grandrt/auriga_public/src/master/}{https://bitbucket.org/grandrt/auriga\_public/src/master/} and that can be used to load and manipulate the data. Some worked examples demonstrating how to use these scripts are provided on the web page.

Auriga is a suite of magnetohydrodynamic cosmological zoom-in simulations of galaxy formation. This consists of several series of simulations: 40 ``level 4'' resolution simulations of halos that fall into the mass range of $[0.5, 2]\times 10^{12}$ $\rm M_{\odot}$ at $z=0$, thus spanning the entire allowed mass range for the Milky Way, and 6 of these are available at ``level 3'' high resolution; 12 simulations of massive dwarf halos in the mass range of $[0.5 ,5]\times 10^{11}$ $\rm M_{\odot}$ at $z=0$ (at both level 4 and level 3 resolution) and a suite of 14 simulations of low-mass dwarf halos in the mass range of $[0.5,5]\times 10^{10}$ $\rm M_{\odot}$ at $z=0$ (at level 3 resolution only). All of these are available with our default magnetohydrodynamics (MHD) model, whereas some are available with hydrodynamics (HD) and dark matter only (DMO).

In the second part of this paper, we have shown that the Auriga simulations offer robust predictions that agree well with many observed galaxy scaling relations over a wide range of mass. In particular, we have demonstrated that the entire set of Auriga simulations compare well with many low-redshift observations, such as the stellar mass-halo mass relation and the scalings of rotation velocity, size, star formation rate, and HI gas fractions as a function of stellar mass. In addition, we have shown how the simulations can be used to connect the different epochs of galaxy formation to help interpret observations: specifically, we showed that the kinematic evolution of star-forming gas traced by H$\alpha$ supports early disc growth followed by ``upside-down'' disc settling of disc formation. 

We hope that this data release will provide the community with an interesting and useful resource to study the formation of star-forming disc galaxies, dwarf galaxies, and their environments. In the future, we plan to make additional newly produced data and associated documentation publicly available via the project website. 

\section*{Data Availability}
All the data presented in this paper is publicly available to download via the Globus platform as described in section \ref{usingdata}.

\section*{Acknowledgements}
RG is supported by an STFC Ernest Rutherford Fellowship (ST/W003643/1). FF is supported by a UKRI Future Leaders Fellowship (grant no. MR/X033740/1). FAG acknowledges support from ANID FONDECYT Regular 1211370, the Max Planck Society through a “Partner Group” grant and ANID Basal Project FB210003. AJ is supported by consolidated grant ST/X001075/1. Part of the simulations of this paper used: the
SuperMUC system at the Leibniz Computing Centre (www.lrz.de), Garching,
under the project PR85JE of the Gauss Centre for Supercomputing e.V. (www.gauss-centre.eu); the Freya computer cluster at the Max Planck Institute for Astrophysics; and the DiRAC@Durham facility managed by the Institute for Computational Cosmology on behalf of the STFC DiRAC HPC Facility (www.dirac.ac.uk). The equipment was funded by BEIS capital funding via STFC capital grants ST/K00042X/1, ST/P002293/1, ST/R002371/1 and ST/S002502/1, Durham University and STFC operations grant ST/R000832/1. DiRAC is part of the National e-Infrastructure.

\bibliographystyle{mnras}
\bibliography{main.bbl}

\end{document}